\documentclass[twocolumn,twocolappendix]{aastex63}

\usepackage{graphbox,graphicx}
\usepackage{dcolumn}
\usepackage{bm}
\usepackage{amssymb}
\usepackage{epstopdf}
\usepackage{amsmath}
\usepackage{amsfonts}
\usepackage{color}
\usepackage{mathrsfs}
\usepackage{xspace}

\definecolor{mediumseagreen}{rgb}{0.24, 0.7, 0.44}

\newcommand{\be}{\begin{equation}}
\newcommand{\ee}{\end{equation}}
\newcommand{\ba}{\begin{eqnarray}}
\newcommand{\ea}{\end{eqnarray}}
\newcommand{\nn}{\nonumber}
\newcommand{\barr}{\begin{array}}
\newcommand{\earr}{\end{array}}

\def\DM{{\rm DM}}
\def\Dg{{\rm DM}_{\rm gal}}
\def\Di{{\rm DM}_{\rm IGM}}
\def\Dh{{\rm DM}_{\rm host}}
\def\Dhh{{\rm DM}_{\rm hh}}
\def\SNR{{\rm SNR}}

\def\Var{\mbox{Var}}
\def\Cov{\mbox{Cov}}
\def\x{{\bf x}}

\def\hC{{\hat C}}

\def\sinc{\mbox{sinc}}
\def\xand{\, \mbox{and} \,}
\def\xor{\, \mbox{or}\,}
\def\pvalue{$p\mkern1.55mu$-value\xspace}
\def\pvalues{$p\mkern1.55mu$-values\xspace}
\def\clfg{\smash{$C_\ell^{fg}$}}
\def\clfgone{\smash{$C_\ell^{fg(1h)}$}}
\def\clfgtwo{\smash{$C_\ell^{fg(2h)}$}}
\def\clgg{\smash{$C_\ell^{gg}$}}
\def\clff{\smash{$C_\ell^{ff}$}}
\def\Pgelz{\,P_{ge}\bigg(\frac{\ell}{\chi(z)},z\bigg)}
\def\smx{\smash{$\SNR_{\rm max}^{(\rm mock)}$}}
\def\sdx{\smash{$\SNR_{\rm max}^{(\rm data)}$}}
\def\pcc{pc\,cm$^{-3}$}

\begin{document}

\title{CHIME/FRB Catalog 1 results: statistical cross-correlations with large-scale structure}
\shorttitle{CHIME/FRB Catalog 1: cross-correlations with LSS}

\author[0000-0001-7694-6650]{Masoud Rafiei-Ravandi}
  \affiliation{Perimeter Institute for Theoretical Physics, 31 Caroline Street N, Waterloo, ON N25 2YL, Canada}
  \affiliation{Department of Physics and Astronomy, University of Waterloo, Waterloo, ON N2L 3G1, Canada}
\author[0000-0002-2088-3125]{Kendrick M.~Smith}
  \affiliation{Perimeter Institute for Theoretical Physics, 31 Caroline Street N, Waterloo, ON N25 2YL, Canada}
\author[0000-0001-7931-0607]{Dongzi Li}
  \affiliation{Cahill Center for Astronomy and Astrophysics, California Institute of Technology, 1216 E. California Boulevard, Pasadena, CA 91125, USA}
\author[0000-0002-4279-6946]{Kiyoshi W.~Masui}
  \affiliation{MIT Kavli Institute for Astrophysics and Space Research, Massachusetts Institute of Technology, 77 Massachusetts Ave., Cambridge, MA 02139, USA}
  \affiliation{Department of Physics, Massachusetts Institute of Technology, 77 Massachusetts Ave., Cambridge, MA 02139, USA}
\author[0000-0003-3059-6223]{Alexander Josephy}
  \affiliation{Department of Physics, McGill University, 3600 rue University, Montr\'eal, QC H3A 2T8, Canada}
  \affiliation{McGill Space Institute, McGill University, 3550 rue University, Montr\'eal, QC H3A 2A7, Canada}
\author[0000-0001-7166-6422]{Matt Dobbs}
  \affiliation{Department of Physics, McGill University, 3600 rue University, Montr\'eal, QC H3A 2T8, Canada}
  \affiliation{McGill Space Institute, McGill University, 3550 rue University, Montr\'eal, QC H3A 2A7, Canada}
\author[0000-0002-1172-0754]{Dustin Lang}
  \affiliation{Perimeter Institute for Theoretical Physics, 31 Caroline Street N, Waterloo, ON N25 2YL, Canada}
  \affiliation{Department of Physics and Astronomy, University of Waterloo, Waterloo, ON N2L 3G1, Canada}
\author[0000-0002-3615-3514]{Mohit Bhardwaj}
  \affiliation{Department of Physics, McGill University, 3600 rue University, Montr\'eal, QC H3A 2T8, Canada}
  \affiliation{McGill Space Institute, McGill University, 3550 rue University, Montr\'eal, QC H3A 2A7, Canada}
\author[0000-0003-3367-1073]{Chitrang Patel}
  \affiliation{Department of Physics, McGill University, 3600 rue University, Montr\'eal, QC H3A 2T8, Canada}
  \affiliation{Dunlap Institute for Astronomy \& Astrophysics, University of Toronto, 50 St.~George Street, Toronto, ON M5S 3H4, Canada}
\author[0000-0003-3772-2798]{Kevin Bandura}
  \affiliation{Lane Department of Computer Science and Electrical Engineering, 1220 Evansdale Drive, PO Box 6109, Morgantown, WV 26506, USA}
  \affiliation{Center for Gravitational Waves and Cosmology, West Virginia University, Chestnut Ridge Research Building, Morgantown, WV 26505, USA}
\author[0000-0002-4064-7883]{Sabrina Berger}
  \affiliation{Department of Physics, McGill University, 3600 rue University, Montr\'eal, QC H3A 2T8, Canada}
  \affiliation{McGill Space Institute, McGill University, 3550 rue University, Montr\'eal, QC H3A 2A7, Canada}
\author[0000-0001-8537-9299]{P.~J.~Boyle}
  \affiliation{Department of Physics, McGill University, 3600 rue University, Montr\'eal, QC H3A 2T8, Canada}
  \affiliation{McGill Space Institute, McGill University, 3550 rue University, Montr\'eal, QC H3A 2A7, Canada}
\author[0000-0002-1800-8233]{Charanjot Brar}
  \affiliation{Department of Physics, McGill University, 3600 rue University, Montr\'eal, QC H3A 2T8, Canada}
\author[0000-0002-2349-3341]{Daniela Breitman}
  \affiliation{Dunlap Institute for Astronomy \& Astrophysics, University of Toronto, 50 St.~George Street, Toronto, ON M5S 3H4, Canada}
  \affiliation{Department of Physics, University of Toronto, 60 St.~George Street, Toronto, ON M5S 1A7, Canada}
  \affiliation{David A.~Dunlap Department of Astronomy \& Astrophysics, University of Toronto, 50 St.~George Street, Toronto, ON M5S 3H4, Canada}
\author[0000-0003-2047-5276]{Tomas Cassanelli}
  \affiliation{Dunlap Institute for Astronomy \& Astrophysics, University of Toronto, 50 St.~George Street, Toronto, ON M5S 3H4, Canada}
  \affiliation{David A.~Dunlap Department of Astronomy \& Astrophysics, University of Toronto, 50 St.~George Street, Toronto, ON M5S 3H4, Canada}
\author[0000-0002-3426-7606]{Pragya Chawla}
  \affiliation{Department of Physics, McGill University, 3600 rue University, Montr\'eal, QC H3A 2T8, Canada}
  \affiliation{McGill Space Institute, McGill University, 3550 rue University, Montr\'eal, QC H3A 2A7, Canada}
\author[0000-0003-4098-5222]{Fengqiu Adam Dong}
  \affiliation{Department of Physics and Astronomy, University of British Columbia, 6224 Agricultural Road, Vancouver, BC V6T 1Z1 Canada}
\author[0000-0001-8384-5049]{Emmanuel Fonseca}
  \affiliation{Department of Physics, McGill University, 3600 rue University, Montr\'eal, QC H3A 2T8, Canada}
  \affiliation{McGill Space Institute, McGill University, 3550 rue University, Montr\'eal, QC H3A 2A7, Canada}
\author[0000-0002-3382-9558]{B.~M.~Gaensler}
  \affiliation{Dunlap Institute for Astronomy \& Astrophysics, University of Toronto, 50 St.~George Street, Toronto, ON M5S 3H4, Canada}
  \affiliation{David A.~Dunlap Department of Astronomy \& Astrophysics, University of Toronto, 50 St.~George Street, Toronto, ON M5S 3H4, Canada}
\author[0000-0001-5553-9167]{Utkarsh Giri}
  \affiliation{Perimeter Institute for Theoretical Physics, 31 Caroline Street N, Waterloo, ON N25 2YL, Canada}
  \affiliation{Department of Physics and Astronomy, University of Waterloo, Waterloo, ON N2L 3G1, Canada}
\author[0000-0003-1884-348X]{Deborah C.~Good}
  \affiliation{Department of Physics and Astronomy, University of British Columbia, 6224 Agricultural Road, Vancouver, BC V6T 1Z1 Canada}
\author[0000-0002-1760-0868]{Mark Halpern}
  \affiliation{Department of Physics and Astronomy, University of British Columbia, 6224 Agricultural Road, Vancouver, BC V6T 1Z1 Canada}
\author[0000-0003-4810-7803]{Jane Kaczmarek}
  \affiliation{Dominion Radio Astrophysical Observatory, Herzberg Research Centre for Astronomy and Astrophysics, National Research Council Canada, PO Box 248, Penticton, BC V2A 6J9, Canada}
\author[0000-0001-9345-0307]{Victoria M.~Kaspi}
  \affiliation{Department of Physics, McGill University, 3600 rue University, Montr\'eal, QC H3A 2T8, Canada}
  \affiliation{McGill Space Institute, McGill University, 3550 rue University, Montr\'eal, QC H3A 2A7, Canada}
\author[0000-0002-4209-7408]{Calvin Leung}
  \affiliation{MIT Kavli Institute for Astrophysics and Space Research, Massachusetts Institute of Technology, 77 Massachusetts Ave., Cambridge, MA 02139, USA}
  \affiliation{Department of Physics, Massachusetts Institute of Technology, 77 Massachusetts Ave., Cambridge, MA 02139, USA}
\author[0000-0001-7453-4273]{Hsiu-Hsien Lin}
  \affiliation{Canadian Institute for Theoretical Astrophysics, 60 St.~George Street, Toronto, ON M5S 3H8, Canada}
\author[0000-0002-0772-9326]{Juan Mena-Parra}
  \affiliation{MIT Kavli Institute for Astrophysics and Space Research, Massachusetts Institute of Technology, 77 Massachusetts Ave., Cambridge, MA 02139, USA}
\author[0000-0001-8845-1225]{B.~W.~Meyers}
  \affiliation{Department of Physics and Astronomy, University of British Columbia, 6224 Agricultural Road, Vancouver, BC V6T 1Z1 Canada}
\author[0000-0002-2551-7554]{D.~Michilli}
  \affiliation{Department of Physics, McGill University, 3600 rue University, Montr\'eal, QC H3A 2T8, Canada}
  \affiliation{McGill Space Institute, McGill University, 3550 rue University, Montr\'eal, QC H3A 2A7, Canada}
\author[0000-0002-3777-7791]{Moritz M\"{u}nchmeyer}
  \affiliation{Department of Physics, University of Wisconsin-Madison, 1150 University Ave., Madison, WI 53706, USA}
\author[0000-0002-3616-5160]{Cherry Ng}
  \affiliation{Dunlap Institute for Astronomy \& Astrophysics, University of Toronto, 50 St.~George Street, Toronto, ON M5S 3H4, Canada}
\author[0000-0002-9822-8008]{Emily Petroff}
  \affiliation{Department of Physics, McGill University, 3600 rue University, Montr\'eal, QC H3A 2T8, Canada}
  \affiliation{McGill Space Institute, McGill University, 3550 rue University, Montr\'eal, QC H3A 2A7, Canada}
  \affiliation{Anton Pannekoek Institute for Astronomy, University of Amsterdam, Science Park 904, 1098 XH Amsterdam, The Netherlands}
  \affiliation{Veni Fellow}
\author[0000-0002-4795-697X]{Ziggy Pleunis}
  \affiliation{Department of Physics, McGill University, 3600 rue University, Montr\'eal, QC H3A 2T8, Canada}
  \affiliation{McGill Space Institute, McGill University, 3550 rue University, Montr\'eal, QC H3A 2A7, Canada}
\author[0000-0003-1842-6096]{Mubdi Rahman}
  \affiliation{Sidrat Research, PO Box 73527 RPO Wychwood, Toronto, ON M6C 4A7, Canada}
\author[0000-0001-5504-229X]{Pranav Sanghavi}
  \affiliation{Lane Department of Computer Science and Electrical Engineering, 1220 Evansdale Drive, PO Box 6109, Morgantown, WV 26506, USA}
  \affiliation{Center for Gravitational Waves and Cosmology, West Virginia University, Chestnut Ridge Research Building, Morgantown, WV 26505, USA}
\author[0000-0002-7374-7119]{Paul Scholz}
  \affiliation{Dunlap Institute for Astronomy \& Astrophysics, University of Toronto, 50 St.~George Street, Toronto, ON M5S 3H4, Canada}
\author[0000-0002-6823-2073]{Kaitlyn Shin}
  \affiliation{MIT Kavli Institute for Astrophysics and Space Research, Massachusetts Institute of Technology, 77 Massachusetts Ave., Cambridge, MA 02139, USA}
  \affiliation{Department of Physics, Massachusetts Institute of Technology, 77 Massachusetts Ave., Cambridge, MA 02139, USA}
\author[0000-0001-9784-8670]{Ingrid H.~Stairs}
  \affiliation{Department of Physics and Astronomy, University of British Columbia, 6224 Agricultural Road, Vancouver, BC V6T 1Z1 Canada}
\author[0000-0003-2548-2926]{Shriharsh P.~Tendulkar}
  \affiliation{National Centre for Radio Astrophysics, Post Bag 3, Ganeshkhind, Pune, 411007, India}
  \affiliation{Department of Astronomy and Astrophysics, Tata Institute of Fundamental Research, Mumbai, 400005, India}
\author[0000-0003-4535-9378]{Keith Vanderlinde}
  \affiliation{Dunlap Institute for Astronomy \& Astrophysics, University of Toronto, 50 St.~George Street, Toronto, ON M5S 3H4, Canada}
  \affiliation{David A.~Dunlap Department of Astronomy \& Astrophysics, University of Toronto, 50 St.~George Street, Toronto, ON M5S 3H4, Canada}
\author[0000-0001-8278-1936]{Andrew Zwaniga}
  \affiliation{Department of Physics, McGill University, 3600 rue University, Montr\'eal, QC H3A 2T8, Canada}
\newcommand{\acks}{
K.W.M. is supported by an NSF grant (2008031).
M.D. is supported by a Killam Fellowship, NSERC Discovery Grant, CIFAR, and by the FRQNT Centre de Recherche en Astrophysique du Qu\'ebec (CRAQ).
M.B. is supported by an FRQNT Doctoral Research Award.
K. B. is supported by an NSF grant (2006548).
P.C. is supported by an FRQNT Doctoral Research Award.
B.M.G. is supported by an NSERC Discovery grant (RGPIN-2015-05948), and by the Canada Research Chairs (CRC) program.
D.C.G is supported by the John I. Watters Research Fellowship.
V.M.K. holds the Lorne Trottier Chair in Astrophysics \& Cosmology, a Distinguished James McGill Professorship and receives support from an NSERC Discovery Grant (RGPIN 228738-13) and Gerhard Herzberg Award, from an R.~Howard Webster Foundation Fellowship from CIFAR, and from the FRQNT CRAQ.
C.L. was supported by the U.S. Department of Defense (DoD) through the National Defense Science \& Engineering Graduate Fellowship (NDSEG) Program.
J.M.-P is a Kavli Fellow.
D.M. is a Banting Fellow.
E.P. acknowledges funding from an NWO Veni Fellowship.
P.S. is a Dunlap Fellow and an NSERC Postdoctoral Fellow.
K.S. is supported by the NSF Graduate Research Fellowship Program.
FRB research at UBC is supported by an NSERC Discovery Grant and by CIFAR.
}

\correspondingauthor{Masoud Rafiei-Ravandi}
\email{mrafieiravandi@perimeterinstitute.ca}

\begin{abstract}
The CHIME/FRB Project has recently released its first catalog of fast radio bursts (FRBs), 
containing 492 unique sources.
We present results from angular cross-correlations of CHIME/FRB sources with
galaxy catalogs.
We find a statistically significant (\pvalue $\sim 10^{-4}$, accounting for look-elsewhere factors)
cross-correlation between CHIME FRBs
and galaxies in the redshift range $0.3 \lesssim z \lesssim 0.5$, in three photometric galaxy
surveys: WISE$\times$SCOS, DESI-BGS, and DESI-LRG.
The level of cross-correlation is consistent with an order-one fraction of the CHIME FRBs being
in the same dark matter halos as survey galaxies in this redshift range.
We find statistical evidence for a population of FRBs with large host dispersion measure ($\sim 400$ \pcc),
and show that this can plausibly arise from gas in large halos ($M \sim 10^{14} M_\odot$),
for FRBs near the halo center ($r \lesssim 100$ kpc).
These results will improve in future CHIME/FRB catalogs, with more FRBs and
better angular resolution.
\end{abstract}


\section{Introduction}
\label{sec:introduction}

Fast radio bursts (FRBs) are millisecond flashes of radio waves whose dispersion is beyond what we
expect from Galactic models along the line of sight. The origin of FRBs is still a mystery,
despite over a decade of observations and theoretical
exploration \citep[see, e.g.][]{Cordes:2019wl,Petroff:2019vo,Platts:2019ws}.
The Canadian Hydrogen Intensity Mapping Experiment~/~Fast Radio
Burst Project \citep[CHIME/FRB;][]{Collaboration:2018aa}
has recently released its first catalog of FRBs containing
492 unique sources \citep{Collaboration:2021wz}, increasing the number of known FRBs by a factor 
$\sim$4.\footnote{For a complete list of known FRBs, see \url{https://www.herta-experiment.org/frbstats}
\citep{Spanakis-Misirlis:2021ta} or the Transient Name Server \citep[TNS,][]{2020TNSAN.160....1P}.}
This unprecedented sample size is a new opportunity for statistical studies of FRBs.

The angular resolution of CHIME/FRB
is not sufficient to associate FRBs with unique host galaxies, except
for some FRBs at very low DM, for example a repeating CHIME FRB associated
with M81 \citep{Bhardwaj:2021aa}.
This appears to put some science questions out of reach, such as determining the
redshift distribution of CHIME FRBs.

However, with large enough catalogs of both FRBs and galaxies, it is
possible to associate FRBs with galaxies statistically, using angular
cross-correlations.
Intuitively, if the angular resolution $\theta_f$ of an FRB experiment
is too large for unique host galaxy associations, there will still be an
excess probability (relative to a random point on the sky) to observe FRBs
within distance $\sim\theta_f$ of a galaxy.
Formally, this corresponds to a cross-correlation between the FRB and
galaxy catalogs, which we will define precisely in~\S\ref{sec:clustering_results}.
By measuring the correlation as a function of galaxy redshift and FRB
dispersion measure (DM) (defined below), the redshift distribution
and related properties of the FRB population can be constrained, even in
the absence of per-object associations.

FRB-galaxy cross-correlations have been proposed in a forecasting context
\citep{McQuinn:2014aa,Masui:2015wu,Shirasaki:2017vd,Madhavacheril:2019vm,
Rafiei-Ravandi:2020aa,Reischke:2021th,Alonso:2021aa,Reischke:2021vx},
and applied to the ASKAP and 2MPZ/HIPASS catalogs by \cite{Li:2019fsg}.
In this paper, we will use machinery developed by \cite{Rafiei-Ravandi:2020aa} 
for modeling the FRB-galaxy cross-correlation, and disentangling it from
propagation effects. This machinery uses the halo model for cosmological large-scale
structure (LSS); for a review see~\cite{COORAY_2002}.

Before summarizing the main results presented here, we recall the
definition of FRB DM.
FRBs are dispersed: the arrival time at radio frequency $\nu$ is
delayed, by an amount proportional to $\nu^{-2}$.
The dispersion is
proportional to the DM, defined as the free electron column density 
along the line of sight:
\be
\DM \equiv \int n_e(x) \, dx \, .
\ee
Since FRBs have not been observed to have spectral lines, FRB redshifts are not directly observable.
However, the DM is a rough proxy for redshift \citep{Macquart:2020aa}.
We write the total DM as the sum of contributions from our Galaxy and halo ($\Dg$),
the IGM ($\Di$), and the FRB host galaxy and halo ($\Dh$):
\be
\DM = \Dg + \Di(z) + \Dh \, .
\ee
The IGM contribution $\Di(z)$ is given by the Macquart relation:
\be
\Di(z) = n_{e,0} \int_0^z dz' \, f_d(z') \frac{1+z'}{H(z')} \, , \label{eq:dm_igm}
\ee
where $f_d(z)$ is the mean electron ionization fraction at redshift $z$,
$n_{e,0}=2.13\times10^{-7}\,{\rm cm}^{-3}$ is the comoving electron density,
and $H(z)$ is the Hubble expansion rate.
If $f_d$ is assumed independent of redshift, then Eq.~(\ref{eq:dm_igm}) has
the following useful approximation:
\be
\Di(z) \approx \Big( 1000 \mbox{ \pcc} \Big) f_d z \, .
\ee
We checked that this approximation is accurate to 6\% for $z \le 3$, assuming
that helium reionization is complete by $z=3$.
By default, we assume $f_d=0.9$, which implies $\Di(z) \approx 900z$ \pcc.

We briefly summarize the main results of the paper.
We find a statistically significant 
correlation between CHIME FRBs and
galaxies in the redshift range $0.3 \lesssim z \lesssim 0.5$.
The correlation is seen in three photometric galaxy surveys: WISE$\times$SCOS,
DESI-BGS, and DESI-LRG (described in~\S\ref{ssec:galaxy_catalogs}).
The statistical significance of the detection in each survey is 
$p \sim (2.7 \times 10^{-5})$, $(3.1 \times 10^{-4})$, and $(4.1 \times 10^{-4})$,
respectively.
These \pvalues account for look-elsewhere effects, in both angular
scale and redshift range.
The observed level of correlation is consistent with an order-one fraction of CHIME FRBs
inhabiting the same dark matter halos as galaxies
in these surveys. CHIME/FRB does not resolve halos, so we cannot
distinguish between FRBs in survey galaxies and FRBs in the
same halos as survey galaxies.

We study the DM dependence of the FRB-galaxy correlation and find
a correlation between high-DM (extragalactic $\DM \ge 785$ \pcc) FRBs and galaxies at $z \sim 0.4$.
This implies the existence of an FRB subpopulation with host DM $\gtrsim 400$ \pcc.
Such large host DMs have not yet been seen in observations that directly
associate FRBs with host galaxies.
To date, 14 FRBs (excluding a Galactic magnetar, see \cite{chimfrb:2020vo,Bochenek_2020})
have been localized to host galaxies, all of which have $\Dh \lesssim 200$ \pcc.
In \S\ref{ssec:dm_dependence}, we explain why these observations are not
in conflict.
We also show that host DMs $\gtrsim 400$ \pcc\ can arise from ionized gas in
large ($M \gtrsim 10^{14} \, M_\odot$) dark matter halos, if FRBs are
located near the halo center ($r \lesssim 100$ kpc).

This paper is structured as follows.
In~\S\ref{sec:data}, we describe the observations and data reduction. 
Clustering results are presented in~\S\ref{sec:clustering_results}
and interpreted in~\S\ref{sec:interpretation}.
We conclude in~\S\ref{sec:discussion}.
Throughout, we adopt a flat $\Lambda$CDM cosmology with
Hubble expansion rate $h=0.67$, 
matter abundance $\Omega_m=0.315$,
baryon abundance $\Omega_b=0.048$, 
initial power spectrum amplitude $A_s=2.10\times10^{-9}$, 
spectral index $n_s=0.965$, 
neutrino mass $\sum_\nu m_\nu=0.06$~eV, 
and CMB temperature $T_{\rm CMB}=2.726$~K\@.
These parameters are consistent with Planck results \citep{Planck_2018}.

\section{Data}
\label{sec:data}

\subsection{FRB catalog}
\label{ssec:frb_catalog}

The first CHIME/FRB catalog is described in \citep{Collaboration:2021wz}.
In order to maximize localization precision and to simplify selection biases, 
we include only a single burst with the highest significance for each repeating FRB in this analysis.
This treats repeating and nonrepeating FRBs as a single population.
In future CHIME/FRB catalogs with more repeaters, it would be interesting to
analyze the two populations separately.
In CHIME/FRB, there is currently no evidence that repeaters and nonrepeaters
have different sky distributions \citep{Collaboration:2021wz}.
We also exclude three sidelobe detections (FRB20190210D, FRB20190125B, FRB20190202B),
leaving a sample of 489 unique sources.
We do not exclude FRBs with {\tt \detokenize{excluded_flag=1}}, indicating an epoch
of low sensitivity, since we expect the localization accuracy of such FRBs to
be similar to the main catalog.

Throughout this paper, {\em all DM values are extragalactic.}
That is, before further processing of the CHIME FRBs, 
we subtract the Galactic contribution $\Dg$ from the observed DM.
The value of $\Dg$ is estimated using the YMW16 \citep{Yao:2017aa} model.
In~\S\ref{ssec:dm_dependence}, we show that using the NE2001
\citep{Cordes:2002tt} model does not affect results qualitatively.
The CHIME/FRB extragalactic DM distribution is shown in Figure~\ref{fig:dndz}.

We do not subtract an estimate of the Milky Way halo DM,
since the halo DM is currently poorly constrained by observations.
The range of allowed values is roughly $10 \lesssim \DM_{\rm halo} \lesssim 100$ \pcc,
and the (dipole-dominated) anisotropy is expected to be 
small~\citep{Prochaska:2019ab,Keating:2020aa}.
The results of this paper are qualitatively unaffected by the
value of $\DM_{\rm halo}$.

The CHIME/FRB pipeline assigns a nominal sky location to each FRB
based on the observed signal-to-noise ratio (SNR) in each of 1024 formed beams.
In the simplest case of an FRB that is detected only in a single formed beam,
the nominal location is the center of the formed beam.
For multibeam detections, the nominal location is roughly a weighted
average of the beam centers \citep{Collaboration:2019tw,Collaboration:2021wz}.
Statistical errors on CHIME/FRB locations are difficult to model,
since they depend on both the details of the CHIME telescope and selection biases
that depend on the underlying FRB population.
We discuss this further in~\S\ref{ssec:pipeline_overview}
and Appendix~\ref{app:sec:localization}.

\subsection{Galaxy catalogs}
\label{ssec:galaxy_catalogs}

\begin{figure*}
\centerline{
        \includegraphics[align=t,width=8.6cm]{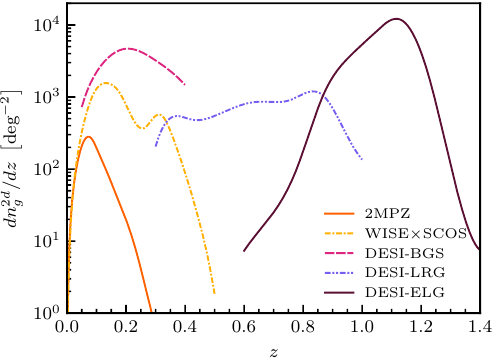}
        \hspace{0.75cm}
        \includegraphics[align=t,width=8.404545454545454cm]{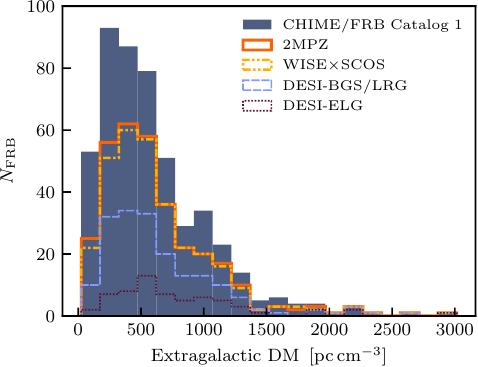}
}
\caption{{\em Left panel:} Redshift distributions for the five galaxy samples
in this paper (\S\ref{ssec:galaxy_catalogs}). 
{\em Right panel:} FRB extragalactic DM distributions for the CHIME/FRB
catalog (solid) and for the subset of the CHIME/FRB catalog that overlaps spatially with
each galaxy survey.}
\label{fig:dndz}
\end{figure*}

On the galaxy side, we have chosen five photometric redshift catalogs:
2MPZ, WISE$\times$SCOS, DESI-BGS, DESI-LRG, and DESI-ELG.
Note that the DESI catalogs are the photometric target samples for
forthcoming spectroscopic DESI surveys with the same names.
Table~\ref{tab:galaxy_catalogs} summarizes key properties of our reduced samples for the
cross-correlation analysis, and the redshift distributions are shown in Figure~\ref{fig:dndz}.

\begin{table}
\begin{center}
\begin{tabular}{@{\hskip 0cm}ccccc@{\hskip 0.01cm}c@{\hskip -0.001cm}}
\hline\hline
    Survey & $f_{\rm sky}$ &
    $[z_{\rm min},z_{\rm max}]$ &
    $z_{\rm med}$ &
    $N_{\rm gal}$ & 
    $N_{\rm FRB}$ \\
\hline
    2MPZ & 0.647 & [0.0,\,0.3] & 0.08 & 670,442 & 323\\
    WISE$\times$SCOS & 0.638 & [0.0,\,0.5] & 0.16 & 6,931,441 & 310\\
    DESI-BGS & 0.118 & [0.05,\,0.4] & 0.22 & 5,304,153 & 183\\
    DESI-LRG & 0.118 & [0.3,\,1.0] & 0.69 & 2,331,043 & 183\\
    DESI-ELG & 0.055 & [0.6,\,1.4] & 1.09 & 5,314,194 & 62\\
    BGS+LRG & 0.118 & [0.05,\,1.0] & 0.28 & 7,690,819 & 183 \\ \hline\hline
\end{tabular}
\end{center}
    \caption{\label{tab:galaxy_catalogs}Galaxy survey parameters: sky fraction $f_{\rm sky}$ (not accounting
for CHIME/FRB coverage), redshift range $[z_{\rm min}, z_{\rm max}]$, median redshift $z_{\rm med}$,
total number of unmasked galaxies $N_{\rm gal}$, and number
of FRBs $N_{\rm FRB}$ overlapping the survey.
The ``BGS+LRG'' catalog is used only in \S\ref{ssec:redshift_dependence}, and consists of all unique 
objects from the DESI-BGS and DESI-LRG catalogs.}
\end{table}

The 2MASS Photometric Redshift (2MPZ) catalog \citep{Bilicki:2013aa} contains $\sim$1~million galaxies with
$z\lesssim 0.3$ (redshift error $\sigma_z\sim0.02$), enabling the construction of a 3D view of LSS at
low redshifts \citep[see, e.g.][]{Alonso:2015aa,Balaguera-Antolinez:2018aa}.  In this work, we use the mask made
by \cite{Alonso:2015aa} for the 2MPZ catalog. Following \cite{Bilicki:2013aa}, 
we discard galaxies whose $K_s$-band magnitude is below the completeness limit $m_{K_s} = 13.9$.

The WISE$\times$SuperCOSMOS photometric redshift catalog \citep[WISE$\times$SCOS,][]{Bilicki:2016aa}
contains $\sim$20~million point sources with $z\lesssim 0.5$ ($\sigma_z\sim0.03$) over 70\% of the sky,
making it a versatile dataset for cross-correlation studies.  In this work, we use a slightly
modified catalog \citep{Krakowski:2016aa}, which includes probabilities 
$(p_{\rm gal}, p_{\rm star}, p_{\rm qso})$ for each object to be a galaxy, star, or quasar, respectively.
We use objects with $p_{\rm gal} \ge 0.9$, which is consistent with the weighted mean purity
of identified galaxies across the $W1$ band \citep{Krakowski:2016aa}.
We use a standard mask\footnote{\url{http://ssa.roe.ac.uk/WISExSCOS.html}}
to remove the Galactic foreground, Magellanic Clouds and bright stars.  Additionally, we mask out
regions that are contaminated visually owing to their proximity to the Galactic plane:
\ba
(|b|\le 20\degr) &\xand& \left((0\degr \le l \le 30\degr) \xor (330\degr \le l \le 360\degr)\right), \nn \\
(|b|\le 18\degr) &\xand& \left((30\degr \le l \le 60\degr) \xor (300\degr \le l \le 330\degr)\right), \nn \\
(|b|\le 17\degr) &\xand& \left(0\degr \le l \le 360\degr\right).
\label{ieq:wise-masks}
\ea

The Dark Energy Spectroscopic Instrument (DESI) Legacy Imaging Surveys~\citep{Dey:2019aa}
were designed to identify galaxies for spectroscopic follow-up.
We use the catalogs from the DR8 release, with photometric redshifts
from~\cite{Zhou:2020ab}.
Following DESI, we consider three samples: the Bright Galaxy Survey (BGS),
the Luminous Red Galaxy (LRG) sample,
and the Emission Line Galaxy (ELG) sample,
corresponding to redshift ranges $0.05 \le z \le 0.4$,
$0.3 \le z \le 1$, and $0.6 \le z \le 1.4$
respectively (Figure~\ref{fig:dndz}).

For each of the three DESI samples, we define survey geometry cuts as follows.
For simplicity, we restrict to the northern part of the survey 
(${\rm Dec} > 32\fdg375$, $b \geq +17\degr$), which contains $\sim$2 times
as many CHIME/FRB sources as the southern part. Note that the northern and southern
DESI surveys are obtained from different telescopes and may have different systematics.
For the DESI-ELG sample,
we impose the additional constraint $b \geq +45\degr$ in order to mitigate systematic depth variations.
We restrict to sky regions that were observed at least twice in each of the $\{g, r, z\}$ bands
\citep{Zhou:2020ab}.
We mask bad pixels, bright stars, large galaxies, and globular clusters using the appropriate DESI
bitmask.\footnote{{\tt MASKBITS} 1, 5--9, and 11--13, defined here: \url{https://www.legacysurvey.org/dr8/bitmasks/}}

In addition to these geometric cuts, we impose per-object cuts on the DESI catalogs
by removing point-like objects ({\tt TYPE=PSF}), and applying the appropriate color
cuts for each of the three surveys.
Color cuts for the BGS, LRG, and ELG catalogs are defined by
\cite{Ruiz-Macias:2020aa},
\cite{Zhou:2020ac},
and \cite{Raichoor:2020aa}
respectively.
For BGS, we include both ``faint'' ($19.5 < r < 20$) and ``bright'' ($r < 19.5$) galaxies
\citep[terminology from][]{Ruiz-Macias:2020aa}.
For BGS and LRG, we exclude objects with poorly 
constrained photometric redshifts ($z_{\rm phot,std} > 0.08$).
Our final BGS, LRG and ELG samples have typical redshift error
$\sigma_z \sim 0.03$, 0.04, and 0.15 respectively.\\

\section{FRB-galaxy correlation results}
\label{sec:clustering_results}

In this section, we describe our pipeline for computing the FRB-galaxy cross
power spectrum. The pipeline consists of mapping sources onto a sky grid and
then computing the spherical harmonic transform and the angular power spectrum.
Error bars are assigned using mock FRB catalogs.

\subsection{Pipeline overview}
\label{ssec:pipeline_overview}

Our central statistic is the angular power spectrum \clfg, a Fourier-space
statistic that measures the level of correlation between the FRB catalog $f$ and
galaxy catalog $g$, as a function of angular wavenumber $\ell$.
Formally, \clfg\ is defined by
\begin{equation}
\big\langle a_{\ell m}^f a_{\ell' m'}^{g*} \big\rangle = C_\ell^{fg} \delta_{\ell\ell'} \delta_{mm'} \, , 
\end{equation}
where $a_{\ell m}^Y$ is the spherical harmonic transform of 
catalog $Y \in \{f,g\}$ (the all-sky analog of the Fourier
transform on the flat sky).
Intuitively, a detection of nonzero \clfg\ at wavenumber $\ell$
corresponds to a pixel-space angular correlation at separation 
$\theta \sim \ell^{-1}$.

The power spectrum \clfg\ is not the only way of
representing a cross-correlation between catalogs as a function of scale.
Another possibility is the correlation function $\zeta(\theta)$,
obtained by counting pairs of objects whose angular separation 
$\theta$ lies in a set of nonoverlapping bins.
This method was used by \cite{Li:2019fsg} to correlate ASKAP FRBs
with nearby galaxies.
The power spectrum \clfg\ and correlation function $\zeta(\theta)$
are related to each other by the Legendre transform
\smash{$\zeta(\theta) = \sum_\ell (2\ell+1)/(4\pi) C_\ell^{fg} P_\ell(\cos\theta)$}.
Therefore, \clfg\ and $\zeta(\theta)$ contain the same information,
and the choice of which one to use is a matter of
convenience.
We have used the power spectrum \clfg, since
it has the property that nonoverlapping $\ell$-bins are nearly
uncorrelated, making it straightforward to infer statistical
significance from plots.

Throughout the paper, it will be useful to have a model
FRB-galaxy power spectrum \clfg\ in mind.
In Figure~\ref{fig:model_clfg}, we show \clfg\ for a galaxy population
at $z \sim 0.4$, calculated using the ``high-$z$'' FRB model 
from~\cite{Rafiei-Ravandi:2020aa}, with median FRB redshift $z=0.76$.
The main features of \clfg\ are as follows:

\begin{figure}[h]
    \centerline{\includegraphics[width=7.928125cm]{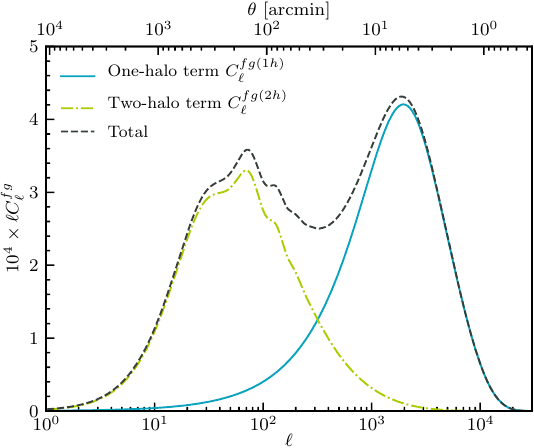}}
    \caption{Model FRB-galaxy power spectrum \clfg\ from \S\ref{ssec:pipeline_overview},
     for a galaxy population near $z\sim 0.37$ and FRB angular resolution $1\arcmin$.
     Note that we have plotted (\smash{$\ell C_\ell^{fg}$}), for consistency with later plots
     in the paper.
     In this and later plots in the paper, the angular scale on the top axis is $\theta=\pi/\ell$,
     and is intended to provide an intuitive mapping between angular multipole $\ell$ and an angular scale.}
\label{fig:model_clfg}
\end{figure}

\begin{itemize}

\item
The leftmost peak at $\ell\sim 10^2$ is the {\em two-halo} term 
\clfgtwo, which arises from FRBs and galaxies in different halos.
The two-halo term does not probe the details of FRB-galaxy associations; 
it arises because FRBs and galaxies both inhabit halos, and halos are
clustered on $\sim 100$\, Mpc scales (the correlation length of the
cosmological density field).

\item
The rightmost peak at $\ell \sim 10^3$ is the {\em one-halo} term
\clfgone, which is sourced by (FRB, galaxy) pairs in the same dark
matter halo.

\item
For completeness, we note that for $\ell \gtrsim 10^4$, there is a ``Poisson'' 
term (not shown in Figure~\ref{fig:model_clfg})
that is sourced by FRBs in catalog galaxies (not elsewhere in the halo).
CHIME/FRB's limited angular resolution suppresses \clfg\ at
high $\ell$, hiding the Poisson term.
Intuitively, this is because CHIME/FRB cannot 
resolve different galaxies in the same dark matter halo.

\end{itemize}

Although the one-halo and two-halo terms look comparable in
Figure~\ref{fig:model_clfg}, the SNR of the one-halo term 
is a few times larger.
In this paper, we do not detect the two-halo term with statistical
significance (see Figure~\ref{fig:clfg_sl_binned}).
Therefore, throughout the paper we will often neglect the two-halo
term, and make the approximation \smash{$C_\ell^{fg} \approx C_\ell^{fg(1h)}$}.

The one-halo term \clfgone\ is constant in $\ell$ for $\ell \lesssim 10^3$,
and suppressed for $\ell \gtrsim 10^3$.
(Note that in Figure~\ref{fig:model_clfg}, we have plotted 
\smash{$\ell C_\ell^{fg}$}, for consistency with later figures in the paper.)
The high-$\ell$ suppression arises
from two effects: (1) statistical errors on FRB positions (the CHIME/FRB ``beam''), and
(2) displacements between FRBs and galaxies in the same dark matter halo.

Within the statistical errors of the \clfg\ measurement in this paper,
both effects can be modeled as Gaussian, i.e. high-$\ell$ suppression 
of the form
$e^{-\ell^2/L^2}$:
\begin{equation}
C_\ell^{fg(1h)} = \alpha e^{-\ell^2/L^2} \, ,  \label{eq:clfg_template}
\end{equation}
where we have omitted the two-halo term since we do not detect it
with statistical significance. 
In principle, the value of $L$ in Eq.~(\ref{eq:clfg_template})
is computable, given models for statistical errors on
CHIME FRB sky locations and FRB/galaxy profiles within 
dark matter halos.
However, FRB halo profiles are currently poorly constrained,
and CHIME FRB location errors are difficult to model, since
they depend on both instrumental selection effects and details
of the FRB population.
In Appendix~\ref{app:sec:localization}, we explore modeling
issues in detail and show that a plausible (but conservative, i.e.~wide)
range of $L$-values is $315 \leq L \leq 1396$.

Summarizing the above discussion, our pipeline works as follows.
We measure the angular power spectrum \clfg\ from the FRB and
galaxy catalogs, and fit the $\ell$-dependence to the template
form \smash{$C_\ell^{fg} = \alpha e^{-\ell^2/L^2}$} in Eq.~(\ref{eq:clfg_template}).
We treat the amplitude $\alpha$ as a free parameter, and
vary the template scale $L$ over the range $315 \le L \le 1396$, to
evaluate the correlation amplitude as a function of scale.

\subsection{Overdensity maps}
\label{ssec:overdensity_maps}

Turning now to implementation,
the first step in our pipeline is to convert the FRB and galaxy catalogs
into ``overdensity'' maps $\delta_f(\x), \delta_g(\x)$, defined by
\be
\delta_Y(\x) = \frac{1}{n_Y^{2d} \, \Omega_{\rm pix}} \Big( N_{Y \in \x} - \overline{N_{Y \in \x}} \Big) \, . \label{eq:deltaX_def}
\ee
Here, $Y \in \{f,g\}$ denotes a catalog, $\x$ denotes an angular pixel,
$N_{Y \in \x}$ denotes the number of catalog objects in pixel $\x$, and
$\overline{N_{Y\in \x}}$ denotes the expected number of catalog objects
in pixel $\x$ due to the survey geometry.
The prefactor $1/(n_Y^{2d} \Omega_{\rm pix})$ is conventional, where 
$n_Y^{2d}$ is the 2D number density and $\Omega_{\rm pix}$ is the pixel area.
For CHIME/FRB, the expected number density $\smash{\overline N_{f\in\x}}$
depends on declination (Dec). The definition~(\ref{eq:deltaX_def}) of $\delta_f(\x)$
weights each pixel $\x$ proportionally to the expected number of FRBs.
This weighting is optimal since the FRB field is Poisson noise dominated
\smash{$(C_\ell^{ff} \approx 1/n_f^{2d})$}.

The difference between a density map and an overdensity map is
the second term $\bar N$ in Eq.~(\ref{eq:deltaX_def}),
which removes spurious density fluctuations due to the survey geometry.
We compute the $\bar N$-term differently for different catalogs as follows.

For the three DESI catalogs, we estimate $\bar N$ using ``randoms''
from the DESI-DR8 release,
i.e.~simulated catalogs that encode the survey geometry, with no
spatial correlations between objects.
We use random catalogs from the DESI-DR8 data release (source density
$n_g^{2d} = 5000$ deg$^{-2}$), and apply the DESI ``geometry'' cuts
from the previous section.

For the other two galaxy surveys (2MPZ and WISE$\times$SCOS), random
catalogs are not readily available, so we represent the survey geometry
by an angular HEALPix~\citep{Gorski:2005aa} mask, and assume uniform
galaxy density outside the mask:
\be
\bar N_{g \in \x} = \left\{ \begin{array}{cl}
  n_g^{2d} \, \Omega_{\rm pix} & \mbox{if $\x$ is unmasked} \\
    0 & \mbox{if $\x$ is masked}
\end{array} \right.
\ee
The mask geometries for 2MPZ and WISE$\times$SCOS were described
previously in~\S\ref{ssec:galaxy_catalogs}.

Finally, for the CHIME/FRB catalog, computing $\bar N$ deserves
some discussion. The CHIME/FRB number density $\bar N$ is inhomogeneous,
peaking near the north celestial pole.
To an excellent approximation, the number density
is azimuthally symmetric in equatorial coordinates, i.e.~independent
of right ascension (RA) at fixed declination, because CHIME
is a cylindrical drift-scan telescope oriented north-south \citep{Collaboration:2021wz}.
Therefore, we make random FRB catalogs that represent $\bar N$
by randomizing RAs of the FRBs in the observed catalog, 
leaving declinations fixed.
When making randoms, we also loop over 1000 copies of the
CHIME/FRB catalog, so that the random catalogs are much larger
than the data catalog (appropriately rescaling $\bar N$ and $n_f^{2d}$
in Eq.~\ref{eq:deltaX_def}).

\begin{figure*}
\centerline{
        \includegraphics[align=t,width=8.6cm]{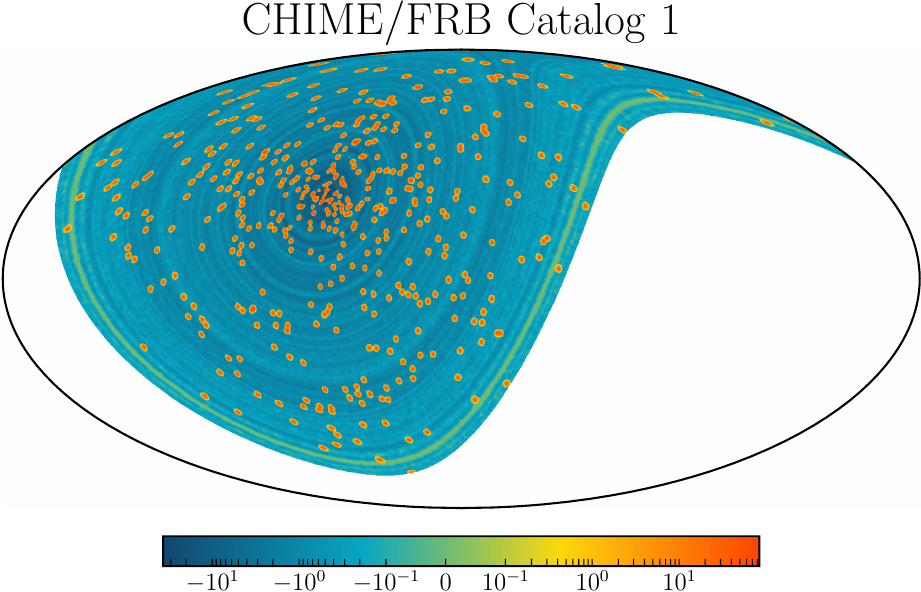}
        \hspace{0.35cm}
        \includegraphics[align=t,width=8.6cm]{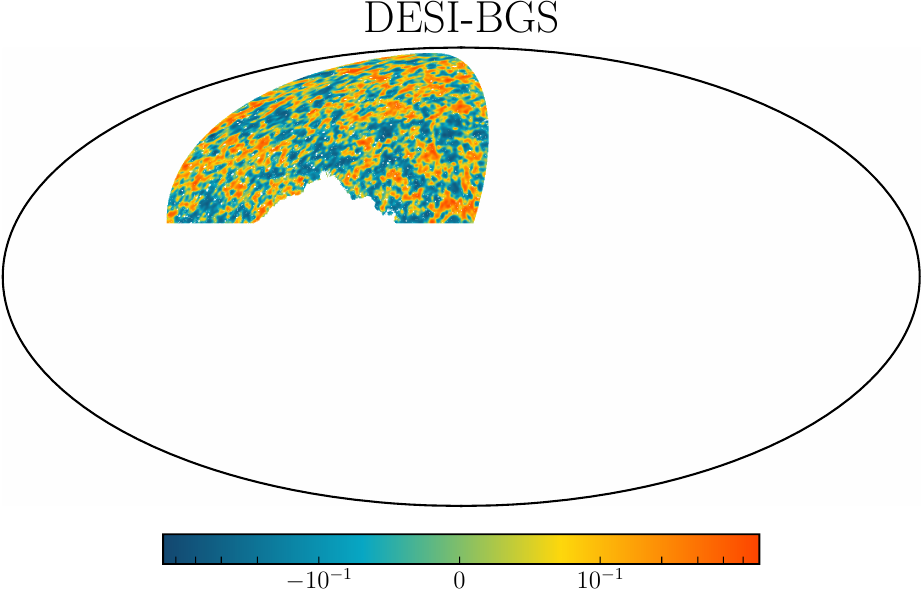}
}
\vspace{0.5cm}
\centerline{
        \includegraphics[align=t,width=8.6cm]{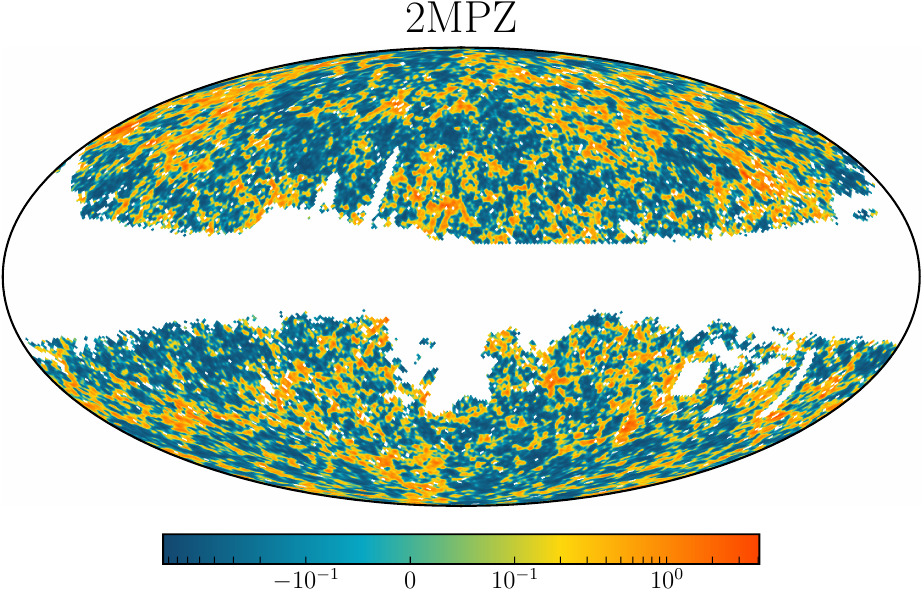}
        \hspace{0.35cm}
        \includegraphics[align=t,width=8.6cm]{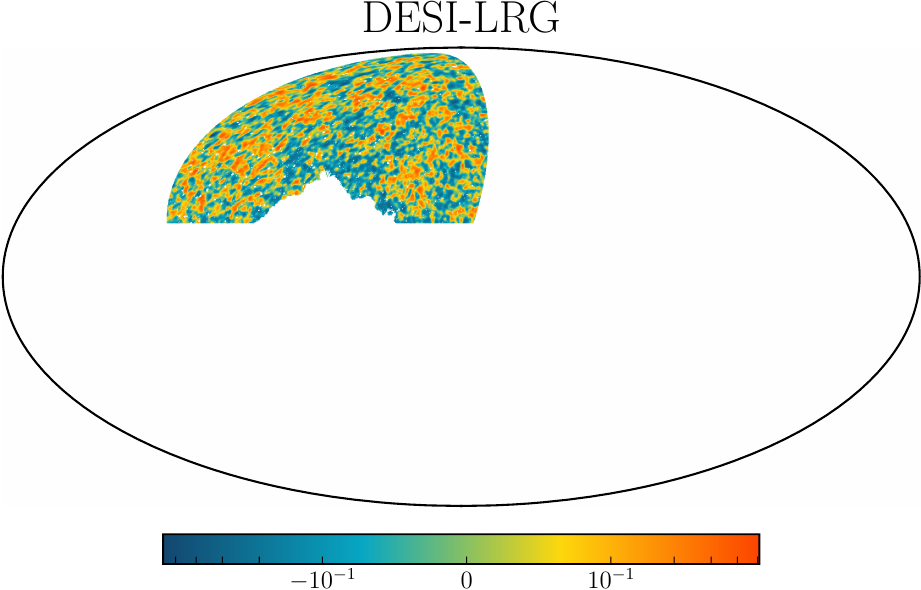}
}
\vspace{0.5cm}
\centerline{
        \includegraphics[align=t,width=8.6cm]{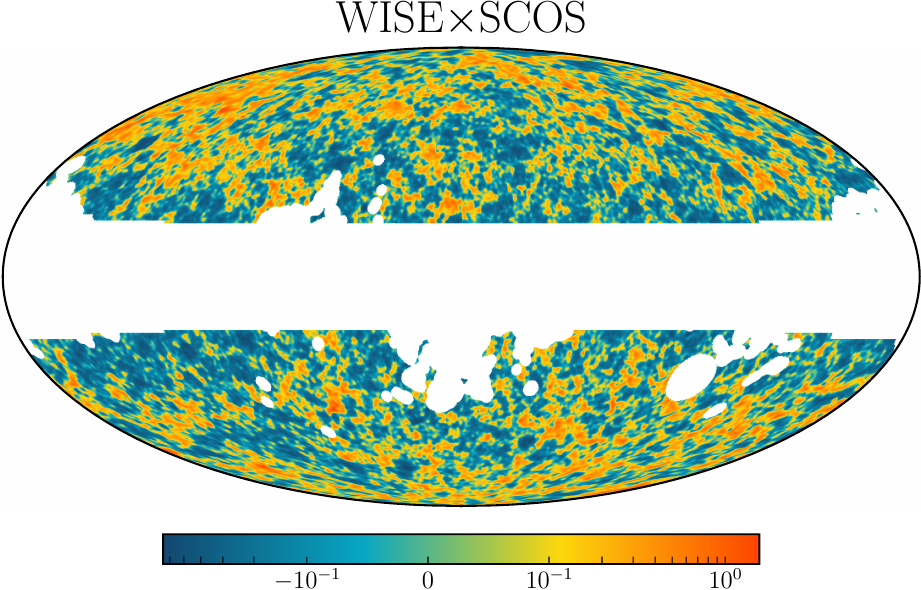}
        \hspace{0.35cm}
        \includegraphics[align=t,width=8.6cm]{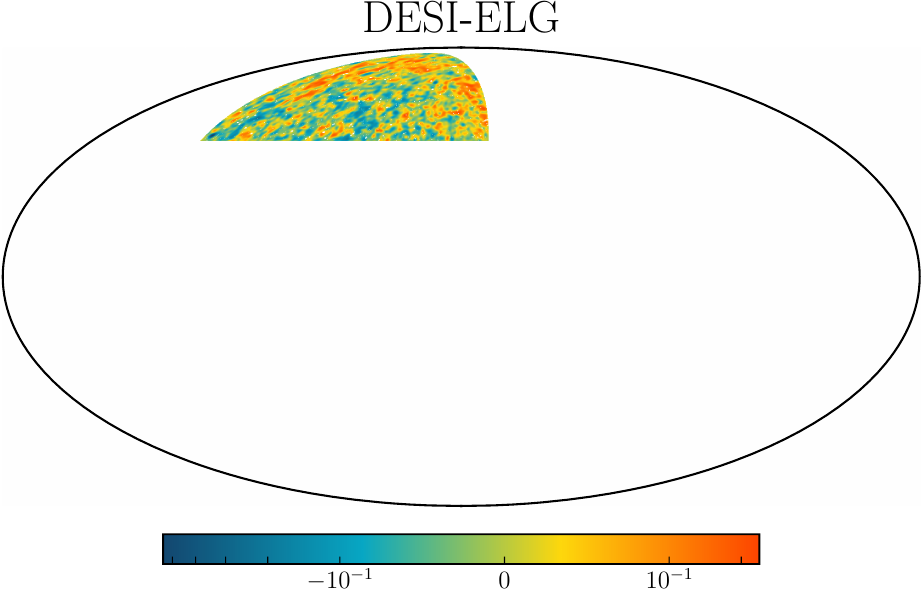}
}
\caption{CHIME/FRB overdensity map $\delta_f(\x)$, and galaxy overdensity maps
 $\delta_g(\x)$ for each galaxy survey. Maps are shown in Mollweide projection,
 centered on $l=180\degr$ in the Galactic coordinate system, after applying
 the angular masks used in the analysis pipeline. To interpret the color
 scale, note that by Eq.~(\ref{eq:deltaX_def}), each object in a pixel contributes 
 $1/(n_Y^{2d} \Omega_{\rm pix})$ to the overdensity $\delta_Y$.}
\label{fig:overdensity_maps}
\end{figure*}

In Figure~\ref{fig:overdensity_maps}, we show overdensity maps $\delta_Y(\x)$ for the 
CHIME/FRB sources and the galaxies.  These maps are useful as visual checks for systematic effects, before
catalogs are cross-correlated.
For example, if the Galactic mask is not conservative enough, the overdensity
map may show visual artifacts with $\delta_g < 0$, since Galactic extinction
will suppress the observed catalog density $N$, relative to $\bar N$.
No visual red flags are seen in either the CHIME/FRB or galaxy maps,
even without a Galactic mask for CHIME/FRB. This is consistent with
\cite{Josephy:2021ts}, who found no evidence for Galactic latitude dependence
in the CHIME/FRB number density after correcting for selection effects.
As described in \S\ref{ssec:galaxy_catalogs}, 
we do apply a Galactic mask in our pipeline, so even if the FRB catalog does
contain low-level biases in the Galactic plane, they should be mitigated.

\subsection{Estimating the power spectrum \clfg}
\label{ssec:clfg}

\begin{figure}[h]
\centerline{\includegraphics[width=8.5cm]{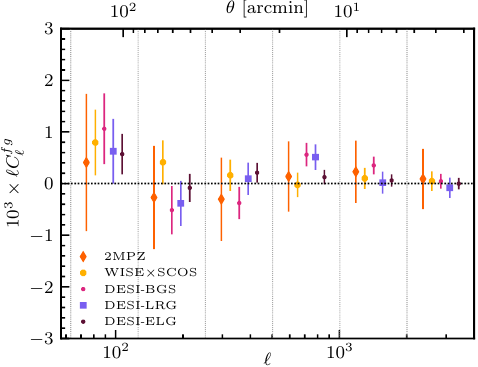}}
\caption{FRB-galaxy cross power spectrum \clfg\ in a set of nonoverlapping $\ell$ bins
  delimited by vertical lines, with $1\sigma$ error bars. Data points are
  shifted slightly from the center of corresponding
  $\ell$ bins for visual clarity.
  Here, we have used all galaxies in the catalogs; if we restrict the redshift
  ranges, then the correlation is more significant (Figure~\ref{fig:clfg_sl_binned}).}
\label{fig:clfg_unbinned}
\end{figure}

We estimate \clfg\ in our pipeline by taking spherical transforms 
of the overdensity maps $\delta_f(\x), \delta_g(\x)$, to get spherical
harmonic coefficients \smash{$a_{\ell m}^f$} and \smash{$a_{\ell m}^g$}.
Then, we estimate the power spectrum \clfg\ as
\be
\hC_\ell^{fg} = \frac{1}{f_{\rm sky}^{fg}} \sum_{m=-\ell}^\ell \frac{1}{2\ell+1} a_{\ell m}^{f*} a_{\ell m}^g \, , \label{eq:hclfg_def}
\ee
where $f_{\rm sky}^{fg}$ is the fractional sky area subtended by the intersection of the FRB and galaxy surveys.
The $f_{\rm sky}^{fg}$ prefactor normalizes the power spectrum estimator to have the correct
normalization on the partial sky.
Throughout the main analysis, we represent overdensities as HEALPix
maps with $1\farcm7$ resolution ($N_{\rm side} = 2048$), and estimate the power spectrum to a maximum
multipole of $\ell_{\rm max}=2000$, corresponding to angular scale $\theta=\pi/\ell_{\rm max}=5\farcm4$.

\begin{figure}
\centerline{\includegraphics[width=8.5cm]{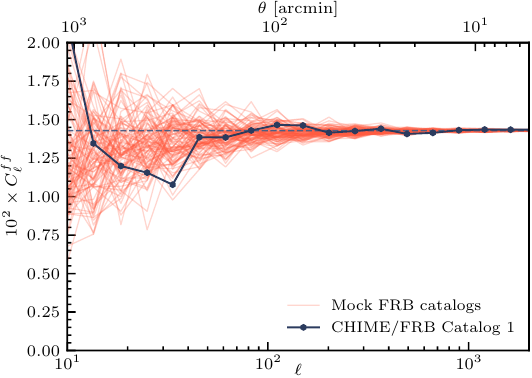}}
\caption{Angular auto power spectrum \clff\ for the CHIME/FRB
    catalog.  Transparent lines represent 100 mock FRB catalogs
    that spatially model the real data.  Throughout the analysis, we assume that
    the power spectrum \clff\ approaches a constant (dashed line) on small scales (high $\ell$).
    Specifically, \smash{$C_\ell^{ff} \approx 1/n_f^{2d}$} for $315 \le \ell \le 1396$.}
\label{fig:clff}
\end{figure}

We assign error bars to the power spectrum \clfg\ using Monte Carlo techniques,
simulating mock FRB catalogs and cross-correlating them with the real
galaxy catalogs.
We simulate mock FRB catalogs by keeping FRB declinations the same as in
the real catalog, but randomizing right ascensions.
This mimics the logic used to construct random FRB catalogs in~\S\ref{ssec:overdensity_maps}.
In fact, the only difference in our pipeline between a ``mock'' and a ``random''
FRB catalog is the number of FRBs: a mock catalog has the same
number of FRBs as the data, whereas a random catalog has a much larger number.
Conceptually, there is another difference between mocks and
randoms: mocks should include any spatial clustering signal present in the real data, whereas 
randoms are unclustered and only represent the survey geometry.  For FRBs,
spatial clustering is small compared to Poisson noise (\clff\ $\approx 1/n_f^{2d}$, 
see Figure~\ref{fig:clff}), so we can make the approximation that clustering is negligible.

\subsection{Statistical significance and look-elsewhere effect}
\label{ssec:statistical_significance}

In Figure~\ref{fig:clfg_unbinned}, we show the angular power spectrum \clfg\
for a set of nonoverlapping $\ell$ bins.
A weak positive FRB-galaxy correlation is seen 
at $500 \lesssim \ell \lesssim 1000$
in some of the galaxy surveys.
In this subsection, we will address the question of whether this correlation is
statistically significant.

As explained in~\S\ref{ssec:pipeline_overview}, we will fit the
FRB-galaxy correlation to the template \smash{$C_\ell^{fg} = \alpha e^{-\ell^2/L^2}$},
treating the amplitude $\alpha$ as a free parameter, and
varying the template scale $L$ over the range $315 \le L \le 1396$.
Let us temporarily assume that $L$ is known in advance.
In this case, an optimal estimator for $\alpha$ is
\be
\hat\alpha_L = \frac{1}{{\mathcal N}_L} \sum_{\ell\ge \ell_{\rm min}} (2\ell+1) \frac{e^{-\ell^2/L^2}}{C_\ell^{gg}} \hC_\ell^{fg} \, , \label{eq:alpha-hat}
\ee
where \smash{$\hC_\ell^{fg}$} was defined in Eq.~(\ref{eq:hclfg_def}),
and the normalization ${\mathcal N}_L$ is defined by
\be
{\mathcal N}_L = \sum_{\ell\ge \ell_{\rm min}} (2\ell+1) \frac{e^{-2\ell^2/L^2}}{C_\ell^{gg}} \, .
\ee
We have included a cutoff at $\ell_{\rm min}=50$ 
to mitigate possible large-scale systematics. This is
a conservative choice, since Figure~\ref{fig:clff} does
not show evidence for systematic power in the auto power
spectrum \clff\ for $\ell \gtrsim 15$.
Eq.~(\ref{eq:alpha-hat}) is derived by noting that
\begin{align}
\Var(\hC_\ell^{fg}) &\propto \frac{C_\ell^{ff} C_\ell^{gg} + (C_\ell^{fg})^2}{2\ell+1} \nn \\
 & \propto \frac{C_\ell^{gg}}{2\ell+1} \, ,
\end{align}
where the first line follows from Wick's theorem,
and the second line follows since \smash{$(C_\ell^{fg})^2 \ll C_\ell^{ff} C_\ell^{gg}$},
and \clff\ is nearly constant in $\ell$.

\begin{figure}[h]
\centerline{\includegraphics[width=8.388646288209607cm]{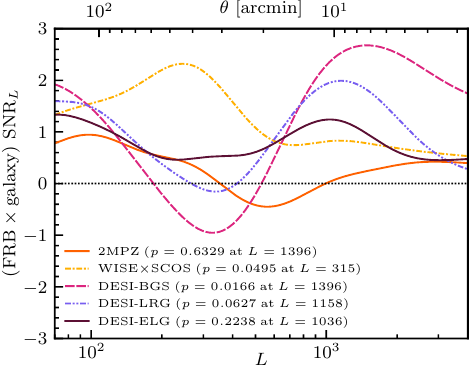}}
    \caption{Quantity $\SNR_L$, defined in Eq.~(\ref{eq:sigma-loc}),
 as a function of template scale $L$. As explained in~\S\ref{ssec:statistical_significance},
 $\SNR_L$ is the statistical significance of the FRB-galaxy correlation in ``sigmas'',
 for a fixed choice of $L$. The \pvalues in the legend are
    bottom-line detection significances after accounting for the look-elsewhere effect in $L$.
  Here, we have used all galaxies in the catalogs; if we restrict the redshift ranges, then
  the detection significance is higher (Figure~\ref{fig:clfg_sl_binned}).}
\label{fig:sigmaL_unbinned}
\end{figure}

We define the quantity
\be
\SNR_L = \frac{\hat\alpha_L}{\Var(\hat\alpha_L)^{1/2}} \, , \label{eq:sigma-loc}
\ee
which is the statistical significance of the \clfg\ detection
in ``sigmas'', for a fixed choice of $L$.
In Figure~\ref{fig:sigmaL_unbinned}, we show the quantity $\SNR_L$, as a function of scale $L$.

We pause for a notational comment: throughout the paper, $\ell$ denotes a
multipole (as in \clfg), and $L$ denotes the template
scale defined in Eq~(\ref{eq:clfg_template}).
The value of $\SNR_L$ (or $\hat\alpha_L$) is obtained by summing \clfg\
over $\ell \lesssim L$, as in Eq.~(\ref{eq:alpha-hat}).
When \clfg\ is computed as a function of $\ell$ (Figure~\ref{fig:clfg_unbinned}),
neighboring $\ell$ bins are nearly uncorrelated, whereas when $\SNR_L$ is computed as
a function of $L$ (Figure~\ref{fig:sigmaL_unbinned}), nearby $L$-values are highly
correlated.

In Figure~\ref{fig:sigmaL_unbinned}, it is seen that $\SNR_L$ can be as large as 2.67,
for a certain choice of $L$ and galaxy survey (namely DESI-BGS at $L=1396$).
However, it would be incorrect to interpret this as a 2.67$\sigma$ detection,
since the value of $L$ has been cherry-picked to maximize the signal.

To quantify statistical significance in a way that accounts for the choice of $L$
(the ``look-elsewhere effect''), we restrict the search to $315 \le L \le 1396$ and
define
\be
\SNR_{\rm max} = \max_{315 \le L \le 1396} \SNR_L \, . \label{eq:sigma_max_def}
\ee
For fixed $L$, $\SNR_L$ is approximately Gaussian distributed, and represents
statistical significance in ``sigmas''. Since $\SNR_{\rm max}$ is obtained by maximizing over
trial $L$-values, $\SNR_{\rm max}$ is non-Gaussian, and we assign statistical significance
by Monte Carlo inference.

In more detail, we compare the ``data'' value of $\SNR_{\rm max}$
(e.g.~$\SNR_{\rm max}=2.67$ for DESI-BGS) to an ensemble of Monte Carlo simulations,
obtained by cross-correlating mock FRB catalogs with the real galaxy
catalog as in~\S\ref{ssec:clfg}.
We assign a \pvalue by computing the fraction of mocks with
$\SNR_{\rm max}^{\rm (mock)} \ge \SNR_{\rm max}^{\rm (data)}$.
We find $p=0.0166$ for DESI-BGS, i.e. evidence for a correlation at 98.34\% CL
after accounting for the look-elsewhere effect in $L$.
The \pvalues for the other galaxy surveys are shown in
Figure~\ref{fig:sigmaL_unbinned}.

Our interpretation is that this level of evidence is intriguing, but
not high enough to be conclusive.
Therefore, we do not interpret the FRB-galaxy correlation in
Figures~\ref{fig:clfg_unbinned} and~\ref{fig:sigmaL_unbinned}
as a detection.
However, in the next subsection we will restrict the redshift
range of the galaxy catalog (accounting for the look-elsewhere
effect in choice of redshift range)
and find a high-significance detection.

\subsection{Redshift dependence}
\label{ssec:redshift_dependence}

\begin{figure*}
\centerline{
        \hspace*{-0.1cm}
        \includegraphics[align=t,width=8.5cm]{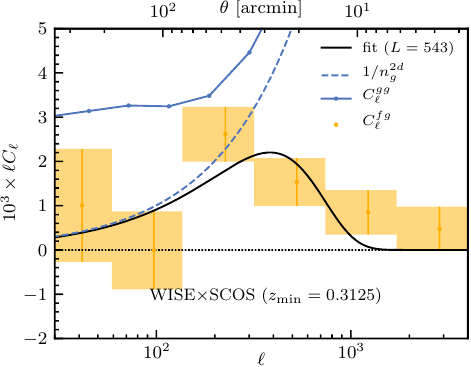}
        \hspace{0.55cm}
        \includegraphics[align=t,width=8.424778761061948cm]{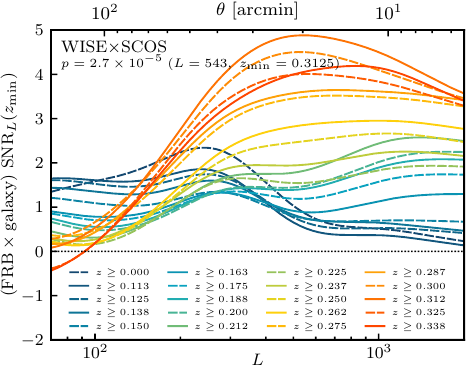}
}
\vspace{0.5cm}
\centerline{
        \hspace*{-0.1cm}
        \includegraphics[align=t,width=8.5cm]{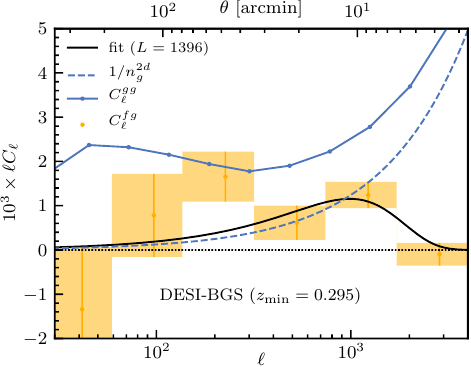}
        \hspace{0.55cm}
        \includegraphics[align=t,width=8.424778761061948cm]{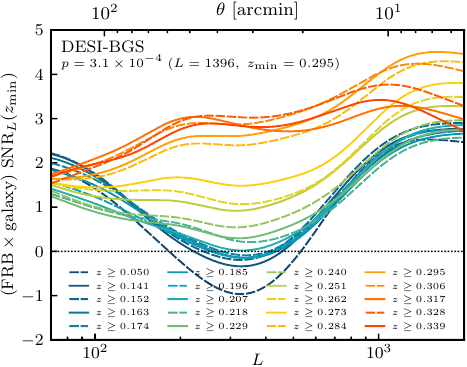}
}
\vspace{0.5cm}
\centerline{
        \hspace*{-0.1cm}
        \includegraphics[align=t,width=8.5cm]{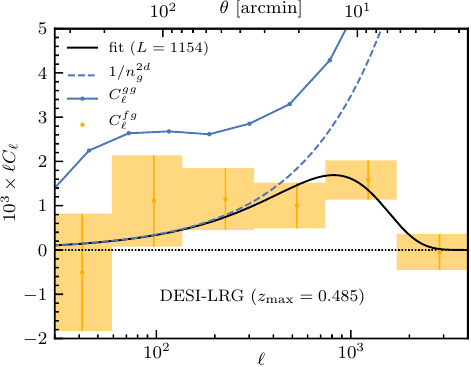}
        \hspace{0.55cm}
        \includegraphics[align=t,width=8.424778761061948cm]{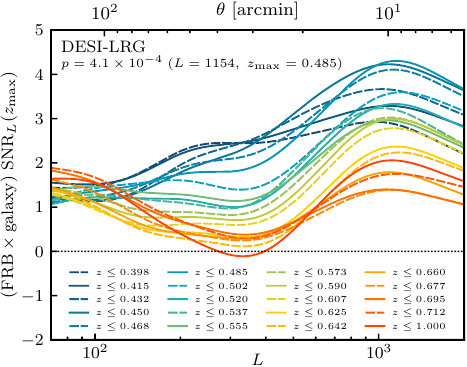}
}
\caption{FRB-galaxy correlation analysis with two parameters: template
 scale $L$ (defined in Eq.~\ref{eq:clfg_template}), and a redshift endpoint 
 (either $z_{\rm min}$ for WISE$\times$SCOS and DESI-BGS, or $z_{\rm max}$ for DESI-LRG).
{\em Left column:} Angular cross power spectrum \clfg\ and auto power spectrum \smash{$C_\ell^{gg}$},
 for the fixed choice of redshift endpoint that maximizes FRB-galaxy correlation.
 The cross power ``fit'' is a best-fit template of the form \smash{$C_\ell^{fg} = \alpha e^{-\ell^2/L^2}$}.
 {\em Right column:} Quantity $\SNR_L$, defined in Eq.~(\ref{eq:sigma-loc}),
 as a function of $L$ and redshift endpoint. 
 As explained in~\S\ref{ssec:statistical_significance}, $\SNR_L$ is
 statistical significance of the FRB-galaxy correlation in ``sigmas'',
 for a fixed choice of $L$ and redshift endpoint.
 The \pvalues in the legend are bottom-line significance after accounting for the
 look-elsewhere effect in these choices (see~\S\ref{ssec:redshift_dependence}).}
\label{fig:clfg_sl_binned}
\end{figure*}

To illustrate our method for studying redshift dependence,
we will use the WISE$\times$SCOS galaxy catalog as a running
example. Suppose we cross-correlate CHIME FRBs with
WISE$\times$SCOS galaxies above some minimum redshift
$z_{\rm min}$, where $z_{\rm min}$ is a free parameter
that will be varied. For each $z_{\rm min}$, we repeat the
analysis of the previous subsection.
The power spectrum \smash{$C_\ell^{fg}(z_{\rm min})$} and
quantity $\SNR_L(z_{\rm min})$ (defined in Eq.~\ref{eq:sigma-loc})
are now functions of two parameters: $z_{\rm min}$ and template
scale $L$.

In the top panels of Figure~\ref{fig:clfg_sl_binned}, we show the power spectrum
\smash{$C_\ell^{fg}(z_{\rm min})$} for the fixed choice of redshift $z_{\rm min}=0.3125$, and
$\SNR_L(z_{\rm min})$ as a function of $L$ and $z_{\rm min}$.
For specific parameter choices, we see a large FRB-galaxy correlation,
e.g.~$\SNR_L(z_{\rm min}) = 4.88$ at $L=543$ and $z_{\rm min} = 0.3125$.
As in the previous subsection, this would imply a 4.88$\sigma$ cross-correlation
for these cherry-picked values of $(L, z_{\rm min})$, but does not account
for the look-elsewhere effect in choosing these values.

To assign statistical significance in a way that accounts for the look-elsewhere effect,
we use the same method as the previous subsection, except that we now scan over
two parameters $(L,z_{\rm min})$ rather than one $(L)$.
Formally, we define
\be
\SNR_{\rm max} = \max_{0 \le z_{\rm min} \le 0.5} \,\, \max_{315 \le L \le 1396} \SNR_L(z_{\rm min}) \, ,
\ee
analogously to Eq.~(\ref{eq:sigma_max_def}) from the previous subsection.
To assign bottom-line statistical significance, we would like
to rank the ``data'' value $\SNR_{\rm max}=4.88$ within a
histogram of $\SNR_{\rm max}$ values obtained by cross-correlating
mock FRB catalogs with the galaxy catalog.
However, with $10^4$ simulations, we find that none of the mock catalogs 
actually exceed $\SNR_{\rm max}=4.88$, so we fit the tail of
the $\SNR_{\rm max}$ distribution to an analytic distribution
(a truncated Gaussian), and compute the \pvalue analytically.
For details of the tail-fitting procedure, see Appendix~\ref{app:tail_fitting}.
We obtain detection significance $p = 2.7 \times 10^{-5}$ for WISE$\times$SCOS
with $z_{\rm min}=0.3125$.
This analysis ``scans'' over minimum redshift $z_{\rm min}$
and scale $L$, and the significance fully accounts for the 
look-elsewhere effect in these parameters.

Similarly, we get $p = 3.1 \times 10^{-4}$ for DESI-BGS
with $z_{\rm min}=0.295$, scanning over $z_{\rm min}$ and $L$.
For DESI-LRG, we use a {\em maximum} redshift $z_{\rm max}$
instead of a minimum redshift $z_{\rm min}$, since DESI-LRG is at
higher redshift than WISE$\times$SCOS or DESI-BGS (Figure~\ref{fig:dndz}).
Scanning over $z_{\rm max}$ and scale $L$, we obtain $p = 4.1 \times 10^{-4}$ 
with $z_{\rm max} = 0.485$ for DESI-LRG.
These results are shown in Figure~\ref{fig:clfg_sl_binned}.

Finally, we find borderline evidence $p=0.0421~(L=1396, z_{\rm max}=0.86)$ 
for a cross-correlation between DESI-ELG galaxies (varying $z_{\rm max}$) and
CHIME FRBs with $\DM \geq 500$ \pcc, where the choice of minimum DM is
fixed. To justify this choice of $\DM_{\rm min}$, note that since host DMs 
must be positive, we do not expect a correlation between DESI-ELG galaxies 
($z_{\rm min} = 0.6$) and CHIME FRBs with $\DM < 500$ \pcc\
(allowing for statistical fluctuations in $\Di$ on the order of 40 \pcc).
We do not find any statistically significant detection with 2MPZ.

These results are consistent with a simple picture in which
the FRB-galaxy correlation mainly comes from galaxies in redshift range
\smash{$0.3 \lesssim z \lesssim 0.5$}.
For WISE$\times$SCOS and DESI-BGS, 
the maximum survey redshifts are 0.5 and 0.4 respectively, and 
we find a strong detection when we impose a
minimum redshift $z_{\rm min} \sim 0.3$.
For DESI-LRG, the minimum survey redshift is 0.3, and we find
a strong detection when we impose a maximum
redshift $z_{\rm max} \sim 0.5$.
The borderline detection in DESI-ELG and nondetection in 2MPZ are also consistent
with this picture, in the sense that these catalogs do not
overlap with the redshift range $0.3 \lesssim z \lesssim 0.5$.

\begin{figure}[h]
    \centerline{\includegraphics[width=8.5cm]{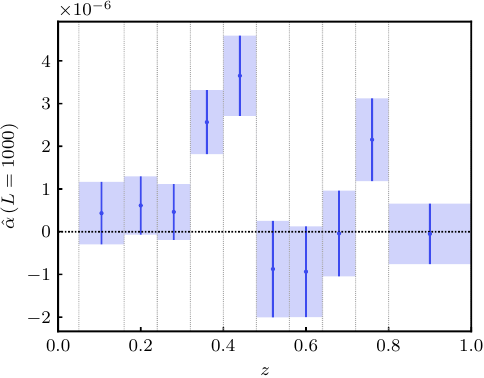}}
    \caption{Redshift dependence of the FRB-galaxy correlation.
 We divide the BGS+LRG catalog into nonoverlapping redshift bins (dotted lines)
 and cross-correlate with CHIME FRBs.
 The quantity $(\hat\alpha_L)_{L=1000}$ on the $y$-axis is a measure
 of the level of cross-correlation, defined in Eq.~(\ref{eq:alpha-hat}).}
\label{fig:alpha_z}
\end{figure}

As a direct way of seeing that the FRB-galaxy correlation
is sourced by redshift range $0.3 \lesssim z \lesssim 0.5$, in Figure~\ref{fig:alpha_z}
we cross-correlate the FRB catalog with the combined BGS+LRG catalog
(Table~\ref{tab:galaxy_catalogs}, bottom row)
in nonoverlapping redshift bins with $0.05 \le z \le 1$.
It is seen that the cross-correlation is driven by redshift range
$0.3 \lesssim z \lesssim 0.5$.
(The bin at $z \sim 0.75$ is nonzero at 2.2$\sigma$, which we interpret 
as borderline statistical significance, since there are 10 bins.)

In Appendix~\ref{app:sec:null_tests}, we examine the robustness 
of these results using null tests and do not find any evidence
for systematic biases.

\section{Interpretation}
\label{sec:interpretation}

So far, we have concentrated on establishing statistical significance
of the FRB-galaxy correlation, in a Monte Carlo simulation pipeline that
accounts for look-elsewhere effects.
In this section, we will interpret the FRB-galaxy correlation, and
explore implications for FRBs.

As explained in~\S\ref{ssec:pipeline_overview},
the output of our pipeline is a constraint on the
coefficient $\alpha$ in the template fit:
\be
C_\ell^{fg} \approx C_\ell^{fg(1h)} \approx \alpha e^{-\ell^2/L^2} \, ,
\ee
where the factor $e^{-\ell^2/L^2}$ is a Gaussian approximation
to the high-$\ell$ suppression due to FRB/galaxy profiles and
the instrumental beam.

At several points in this section, we will want to compare
our FRB-galaxy correlation results to a model for \clfgone.
To do this, we intepret the low-$\ell$ limit of the model as a
prediction for the coefficient $\alpha$ above.
Formally, we define
\be
\alpha \equiv \lim_{\ell\rightarrow 0} C_\ell^{fg(1h)}  \label{eq:alpha_lim0}
\ee
and compare this model prediction for $\alpha$ to the value of $(\hat\alpha_L)_{L=1000}$,
where the estimator $\hat\alpha_L$ was defined in Eq.~(\ref{eq:alpha-hat}).
For simplicity we will fix $L=1000$, since this gives a high-significance
detection of the FRB-galaxy correlation in all three galaxy surveys
(see Figure~\ref{fig:clfg_sl_binned}).

\subsection{Link counting}
\label{ssec:amplitude}

In this subsection, we will interpret the amplitude of the FRB-galaxy
correlation \clfg\ in an intuitive way.
First, we fix a galaxy catalog and redshift range.
As a definition, we say that an FRB is {\em linked} to a galaxy if
they are in the same dark matter halo.
For each FRB $f$, we define the {\em link count} $\eta_f$ by
\begin{equation}
\eta_f = \mbox{number of survey galaxies linked to FRB $f$.}
\end{equation}
Given an FRB catalog, we define the {\em mean link count}
$\eta$:
\begin{equation}
\eta = \big\langle \eta_f \big\rangle \, ,
\end{equation}
where the expectation value $\langle \cdot \rangle$
is taken over FRBs in the catalog.

To connect these definitions with our FRB-galaxy
correlation results, we note that:
\be
\alpha = \lim_{\ell\rightarrow 0} C_\ell^{fg(1h)} = \frac{\eta}{n_g^{2d}} \, , \label{eq:alpha_eta}
\ee
where the first equality is Eq.~(\ref{eq:alpha_lim0}),
and the second equality follows from a short halo
model calculation \citep{Rafiei-Ravandi:2020aa}.
That is, the amplitude $\alpha$ of the FRB-galaxy correlation (in the one-halo regime)
is equivalent to a measurement of the mean link count $\eta$.
This provides a more intuitive interpretation of the amplitude.

In each row of Table~\ref{tab:eta}, we specify
a choice of galaxy catalog and redshift range.
The redshift ranges have been chosen to maximize 
\clfg, as in~\S\ref{ssec:statistical_significance}.
In the third column, we give the constraint on $\alpha$
obtained from the estimator $\hat\alpha_L$ at $L=1000$.
In the last column, we have translated this constraint
of $\alpha$ to a constraint on $\eta$, using Eq.~(\ref{eq:alpha_eta}).

Taken together, the $\eta$ measurements in Table~\ref{tab:eta}
show that the CHIME/FRB catalog has mean link counts of order unity
with galaxies in the range $0.3 \lesssim z \lesssim 0.5$.
The precise value of $\eta$ depends on the specific galaxy survey
considered.
Note that different galaxy surveys will have different values of
$\eta$, since the number of galaxies per halo (and to some extent the
population of halos that is sampled) will be different.

Since FRBs outside the redshift range of the galaxy catalog
do not contribute to $\eta$,
we write $\eta = p \tilde\eta$, where $p$ is the probability
that an FRB is in the catalog redshift range
and $\tilde\eta$ is the mean link count of FRBs that are
in the catalog redshift range.

For the galaxy surveys considered here, we expect $\tilde\eta$ to
be of order unity, since dark matter halos rarely contain more than a few 
catalog galaxies.
To justify this statement, we note that \clgg\ is $\sim 2$ times larger
than the Poisson noise $1/n_g^{2d}$ in the one-halo regime 
(see Figure~\ref{fig:clfg_sl_binned}).
By a link counting argument similar to Eq.~(\ref{eq:alpha_eta}), this
implies that $\langle N_g^2 \rangle \sim 2 \langle N_g \rangle$, where
$N_g$ is the number of galaxies in a halo, and the expectation values
are taken over halos.

Since $\eta = p \tilde\eta$ is of order unity (by Table~\ref{tab:eta}),
and $\tilde\eta$ is of order unity (by the argument in the previous paragraph),
we conclude that $p$ is of order unity. That is,
{\em an order-one fraction of CHIME FRBs are in the redshift range $0.3 \lesssim z \lesssim 0.5$}.

We have phrased this conclusion as a qualitative statement (``order-one fraction'')
since it is difficult to assign a precise upper bound to $\tilde\eta$.
More generally, it is difficult to infer the FRB redshift distribution $(dn_f^{2d}/dz)$ from the
FRB-galaxy correlation in the one-halo regime, since the level of correlation is proportional to
$\tilde\eta (dn_f^{2d}/dz)$, with no obvious way of disentangling the two factors.
Future CHIME/FRB catalogs should contain enough FRBs to detect the FRB-galaxy
correlation on two-halo scales ($\ell \sim 100$)~\citep{Rafiei-Ravandi:2020aa},
which will help break the degeneracy and measure $(dn_f^{2d}/dz)$ and $\tilde\eta$ separately.

\begin{table*}
\begin{tabular}{@{\hskip 0.4cm}c@{\hskip 0.75cm}c@{\hskip 0.75cm}c@{\hskip 0.75cm}c@{\hskip 0.75cm}c@{\hskip 0.35cm}}
\hline\hline
Survey 
   & $[z_{\rm min}, z_{\rm max}]$ 
   & $(\alpha_L)_{L=1000}$ 
   & $n_g^{2d}$ [sr$^{-1}$] 
   & $\eta$ \\ \hline
WISE$\times$SCOS 
   & [0.3125, 0.5]
   & $(4.35 \pm 0.97) \times 10^{-6}$ 
   & $9.92 \times 10^4$ 
   & $0.432 \pm 0.096$ \\
DESI-BGS 
   & [0.295, 0.4]
   & $(2.69 \pm 0.67) \times 10^{-6}$ 
   & $7.94 \times 10^5$ 
   & $2.13 \pm 0.53$ \\
DESI-LRG 
   & [0.3, 0.485]
   & $(3.94 \pm 0.93) \times 10^{-6}$ 
   & $2.83 \times 10^5$ 
   & $1.11 \pm 0.26$ \\  \hline\hline
\end{tabular}
\caption{\label{tab:eta}Clustering analysis in~\S\ref{ssec:amplitude}.
The FRB-galaxy clustering statistic $\alpha_L$
(Eq.~\ref{eq:alpha-hat}) can be translated to a constraint
on $\eta$, the average number of survey galaxies in the same 
dark matter halo as a CHIME/FRB source (see text for details).
The statistical error on $\eta$ in each row is roughly proportional to $n_g^{2d}$.}
\end{table*}

\subsection{DM dependence}
\label{ssec:dm_dependence}

\begin{figure}[h]
    \centerline{\includegraphics[width=8.5cm]{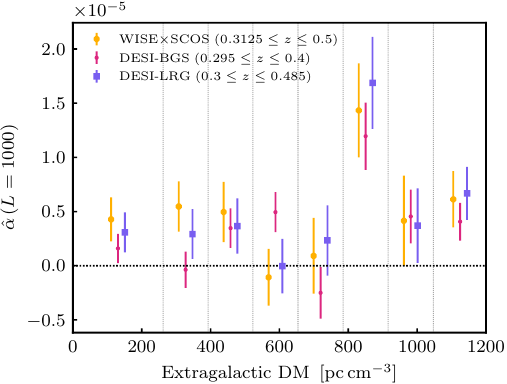}}
\caption{DM dependence of the FRB-galaxy correlation.
We divide the CHIME/FRB catalog into DM bins (delimited by vertical lines)
after subtracting the YMW16 estimate of the Galactic DM and cross-correlate
each DM bin with the galaxy catalogs.
The last DM bin extends to $\DM = 3020$ \pcc.
For each galaxy survey, we use the same redshift range (see legend)
as in the left panel of Figure~\ref{fig:clfg_sl_binned}.
The quantity $(\hat\alpha_L)_{L=1000}$ on the $y$-axis is defined in
Eq.~(\ref{eq:alpha-hat}) and measures the level of FRB-galaxy correlation.
This quantity is a per-object statistic that is derived from \clfg.
Hence, it does not necessarily follow number density variations in Figure~\ref{fig:dndz}.}
\label{fig:alpha_d}
\end{figure}

In Figure~\ref{fig:alpha_d}, we divide the FRB catalog into 
extragalactic DM bins and explore the DM dependence of 
the FRB-galaxy cross-correlation.

A striking feature in Figure~\ref{fig:alpha_d} is the
nonzero correlation in the three highest-DM bins, corresponding
to extragalactic DM $\ge 785$ \pcc.\footnote{A technical
 comment here: for some DM bins in Figure~\ref{fig:alpha_d}, 
 the large values of \clfg\ lead to link counts $\eta$ that are
 a few times larger than the link counts reported in Table~\ref{tab:eta}
 for the whole catalog,
 although statistical errors are large. However, the correlation
 coefficient between the FRB and galaxy fields is never larger than 1.
 In all cases, the field-level correlation \smash{$C_\ell^{fg} / (C_\ell^{ff} C_\ell^{gg})^{1/2}$}
 is of order 0.01 or smaller.}
For reference, the last three bins represent 7\%, 6\%, and 15\% 
of the CHIME/FRB catalog, respectively.
At the redshift of the galaxy surveys ($z \sim 0.4$), the IGM
contribution to the DM is $\Di(z) \sim 360$ \pcc.
Therefore, the observed FRB-galaxy correlation at $\DM \ge 785$ \pcc\
is evidence for a subpopulation of FRBs with
host DMs of order $\Dh \sim 400$ \pcc.

This may appear to be in tension with recent direct associations between FRBs and host galaxies,
which have typically been studied only for lower-redshift FRBs.
At the time of this writing, 14 FRBs have been localized to host 
galaxies,\footnote{\url{https://frbhosts.org/\#explore}}
all of which have $\Dh \lesssim 200$ \pcc\
\citep{Spitler:2016tw,Bassa:2017uf,Chatterjee:2017aa,Kokubo:2017uy,Tendulkar:2017aa, 
Bannister:2019aa,Prochaska:2019aa,Ravi:2019aa,Chittidi:2020aa,Heintz:2020aa,Law:2020aa,
Macquart:2020aa,Mannings:2021ws,Marcote:2020ts,Simha:2020ug,Bhandari:2020aa,Bhandari:2020uj,
Collaboration:2020vl,Bhardwaj:2021aa,James:2021aa}.
The rest of this section is devoted to interpreting this result further.

In Figure~\ref{fig:alpha_d}, the DM bin at $785 < \DM < 916$ \pcc\
is an outlier, suggesting a narrow feature in the DM dependence
of the FRB-galaxy correlation.
Given the error bars, it is difficult to say with statistical significance
whether the apparent narrowness is real, or whether the true DM dependence
is slowly varying.
A crucial point here is that the three galaxy catalogs are highly correlated
spatially (after restricting to the appropriate redshift ranges), which implies
that the three measurements in Figure~\ref{fig:alpha_d} have highly correlated
statistical errors.
Future CHIME/FRB catalogs will have smaller error bars and can
statistically distinguish a narrow feature from slowly varying
DM dependence.

As a check, we remade Figure~\ref{fig:alpha_d}
using the NE2001 \citep{Cordes:2002tt} model for Galactic DM,
instead of the YMW16 model.
The effect of this change is small compared to the statistical
errors in Figure~\ref{fig:alpha_d}.

\begin{figure*}
\centerline{\includegraphics[width=17.8cm]{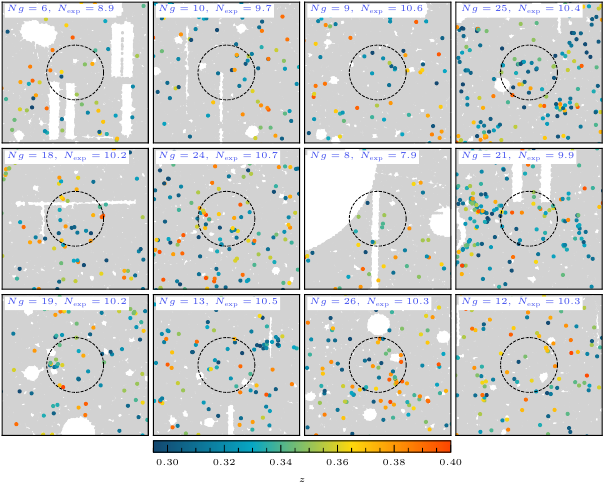}}
    \caption{Visual representation of the cross-correlation between FRBs with
    $785 < \DM < 916$ \pcc, and DESI-BGS galaxies.
    There are 12 FRBs in this DM range in the DESI footprint.
    For each such FRB, we plot the DESI-BGS
    galaxies in the redshift range $0.295 < z < 0.4$ in the vicinity of the FRB.
    We color-code galaxies by redshift, but note that redshift errors are comparable 
    $(\sigma_z \sim 0.03$) to the redshift range shown.
    The gray points are objects in the DESI random catalog, to give a sense for the DESI mask geometry.
    The dashed circles are centered at FRBs, with radius $\Theta=7\arcmin$ (see~\S\ref{ssec:dm_dependence}).
    The value of $N_g$ in the upper left is the observed number of galaxies in the circle.
    The value of $N_{\rm exp}$ is the expected number of galaxies in the circle, inferred from randoms.
    The FRB-galaxy correlation appears as a statistical preference for $N_g > N_{\rm exp}$.}
\label{fig:gallery_bgs}
\end{figure*}

We also performed the following visual check.
The outlier bin with $785 < \DM < 916$ \pcc\ in Figure~\ref{fig:alpha_d} only
contains 12 FRBs in the DESI footprint.
In Figure~\ref{fig:gallery_bgs} we show the DESI-BGS galaxies 
in the vicinity of each FRB.
The large FRB-galaxy correlation can be seen visually as an excess of galaxies
(relative to random catalogs) within $7\arcmin$ of an FRB.\footnote{The scale
 $\Theta \equiv 7\arcmin$ was obtained as $\Theta=\sqrt{8}/L$, where $L=1396$ is the
 template scale where the DESI-BGS cross-correlation peaks in 
 Figure~\ref{fig:clfg_sl_binned}. The factor $\sqrt{8}$ was derived
 by matching the variance $(\Theta^2/2)$ of a radius-$\Theta$ top hat
 to the variance $(4/L^2)$ of a Gaussian beam $e^{-\theta^2L^2/4}$.\label{ft:theta}}
None of the individual FRBs in Figure~\ref{fig:alpha_d} give a statistically
significant cross-correlation on its own, but the total FRB-galaxy correlation
is significant at the 3$\sigma$--4$\sigma$ level.
(We caution the reader that the galaxy counts in Figure~\ref{fig:gallery_bgs}
do not obey Poisson statistics, since the galaxies are clustered.)
There are no visual red flags in Figure~\ref{fig:gallery_bgs}, such as a
single FRB that gives an implausibly large contribution to the cross-correlation.

Finally, we address the question of whether the high-DM signal in
Figure~\ref{fig:alpha_d} is consistent with direct host associations.
Consider the following two statements, in the context of FRB surveys
with the CHIME/FRB sensitivity:
\begin{enumerate}
\item A random FRB with extragalactic $\DM \ge 785$ \pcc\
 has an order-one probability of having redshift $z \sim 0.4$
 (implying $\Dh \gtrsim 400$ \pcc).
\item A random FRB at redshift $z \sim 0.4$
 has an order-one probability of having extragalactic $\DM \ge 785$ \pcc.
\end{enumerate}
The high-DM signal in Figure~\ref{fig:alpha_d} implies statement 1,
but not statement 2.
We will now argue that statement 1 is actually
consistent with direct associations.

The key point is that there are few direct associations at high DM.
Out of the 14 direct associations to date, only one has extragalactic DM $\ge 785$ \pcc:
an FRB with YMW16-subtracted DM 850 \pcc\ at $z=0.6$ \citep{Law:2020aa}.
Based on this one high-DM event, one cannot rule out statement 1 above
(note that statement 2 would clearly be inconsistent with direct associations).

Therefore, there is no inconsistency between the high-DM
FRB-galaxy correlation in Figure~\ref{fig:alpha_d}, and direct
FRB host associations to date.
The number of direct associations is rapidly growing,
and we predict that FRBs with extragalactic DM $\ge 785$ \pcc\ at $z \sim 0.4$
will be found in direct associations soon
(see~\S\ref{sec:discussion} for more discussion).

One final comment: we have presented statistical evidence that statement 1 is true
in CHIME/FRB, but statement 1 depends to some extent on the selection function of the
FRB survey. In particular, future surveys that are sensitive to fainter sources may
detect larger numbers of high-redshift FRBs. In this scenario, it is possible that 
FRBs with extragalactic DM $\ge 785$ \pcc\ will mostly come from $z \sim 0.8$, as
expected from the Macquart relation.

\subsection{Host halo DMs}
\label{ssec:host_halo_dms}

In the previous subsection, we found statistical
evidence for a population of FRBs at $z \sim 0.4$ with
$\DM \gtrsim 400$ \pcc. In this section we will propose
a possible mechanism for generating such large host DMs.
Note that for a Galactic pulsar, a DM of order 400 \pcc\ would be unsurprising,
but pulsar sight lines lie preferentially in the Galactic disk (boosting the DM),
whereas FRBs are observed from a random direction.

Bright galaxies in cosmological surveys are usually found in
large dark matter halos \citep{Wechsler_2018}.
Therefore, FRBs that correlate with such galaxies may have
large host DMs, due to DM contributions from gas in the host 
halos.
We refer to such a contribution as the {\em host halo DM} $\Dhh$,
since the term ``halo DM'' is often used to refer to the contribution
from the Milky Way halo.

\begin{figure}[h]
\centerline{\includegraphics[width=8.5cm]{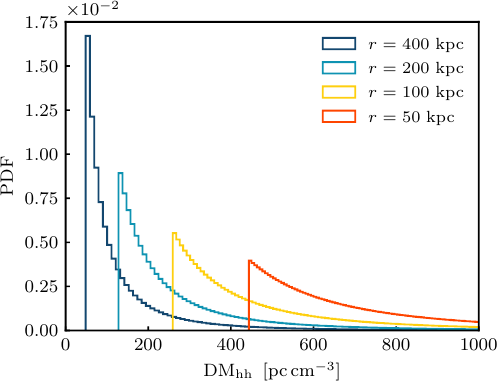}}
\caption{Host halo DM distributions for FRBs in a halo of mass 
$M = 10^{14} M_\odot$. The host halo DM is determined by two
parameters: the distance $r$ between the FRB and halo center,
and viewing angle $\theta$. Each histogram corresponds to one
choice of $r$, with $10^5$ values of $\theta$.
The halo gas profile is the ``ICM'' model
from~\cite{Prochaska:2019ab}.}
\label{fig:Dhh}
\end{figure}

Can host halo DMs plausibly be of order $\Dhh \gtrsim 400$ \pcc?
To answer this question, in Figure~\ref{fig:Dhh}, we show $\Dhh$ 
histograms for simulated FRBs in a halo of mass $M = 10^{14} M_\odot$.
The halo gas profile is the intracluster medium (ICM) model from~\cite{Prochaska:2019ab},
based on X-ray observations from
\cite{Vikhlinin_2006}\footnote{To calculate the host halo DM
$\Dhh = \int dr \, n_e(r)$, we used a slightly modified version
of the {\tt FRB} software (\url{github.com/FRBs/FRB})
by Prochaska et al. We thank the authors for making their software
public.}.
It is seen that FRBs near the centers
($r \lesssim 100$ kpc) of large ($M \sim 10^{14} M_\odot$) 
halos can have host halo DMs $\Dhh \gtrsim 400$ \pcc.

Thus, the high-DM signal in Figure~\ref{fig:alpha_d}
is plausibly explained by a small subpopulation of FRBs at
redshift $0.3 \lesssim z \lesssim 0.5$ near the centers of
large halos. Such a subpopulation could have
$\Dh \gtrsim 400$ \pcc, and strongly correlate with galaxies,
since bright galaxies are often in high-mass halos.

This mechanism is a proof of concept to show that $\Dhh \gtrsim 400$ \pcc\
is plausible in some halo gas models.
Other mechanisms may also be possible, such as augmentation by intervening
foreground galaxies \citep{James:2021ab}. We emphasize that the statistical
evidence for a population of FRBs with $\Dh \gtrsim 400$ \pcc, presented in
the previous subsection, does not depend on the assumption of a particular model
or mechanism.

\subsection{Propagation effects}
\label{ssec:propagation_effects}

So far, we have assumed that the observed FRB-galaxy
correlation is owing to spatial correlations between
the FRB and galaxy populations.
In this subsection, we will explore the alternate hypothesis
that host DMs are always small (say $\Dh \sim 70$ \pcc), and
that propagation effects are responsible for the observed
correlation between $z\sim 0.4$ galaxies and high-DM
FRBs.

``Propagation effects'' is a catch-all term for
what happens to radio waves during their voyage
from source and observer due to intervening plasma.
For example, dispersion, scattering, and plasma lensing
are all propagation effects.
Propagation effects can produce an apparent correlation between
low-redshift galaxies and high-redshift FRBs, even when the
underlying populations are not spatially correlated.

For example, low-redshift galaxies are spatially correlated
with free electrons, which contribute to the DM of background FRBs.
The DM contribution can either increase or decrease the probability
of detecting a background FRB, depending on the selection
function of the instrument.
This effect can produce an apparent correlation or 
anticorrelation between low-$z$ galaxies and high-$z$
FRBs, in the absence of any spatial correlation between
the galaxy and FRB populations.

Here, we will calculate contributions to 
\clfg\ from propagation effects, using formalism from
\cite{Rafiei-Ravandi:2020aa}.
We will use a fiducial model in which host DMs are small
($\Dh \sim 70$ \pcc), implying negligible spatial correlation
between $z\sim 0.4$ galaxies and high-DM FRBs.
This is because we are interested in exploring the hypothesis
that propagation effects (not large host DMs) are entirely
responsible for the observed DM dependence in Figure~\ref{fig:alpha_d}.
We describe the fiducial model in the next few paragraphs.

\begin{figure}
\centerline{\includegraphics[width=8.251046025104602cm]{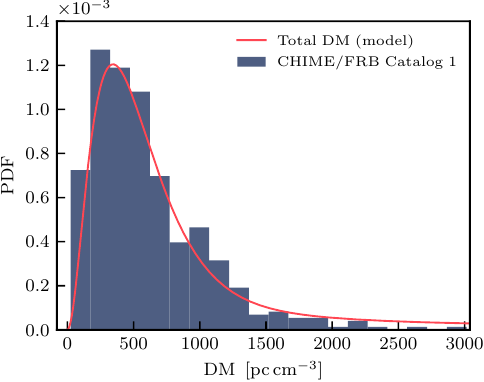}}
\caption{DM distribution (solid curve) for the fiducial FRB model used
 to study propagation effects in~\S\ref{ssec:propagation_effects},
 with the CHIME/FRB DM distribution shown for comparison (histogram).
 In this model, host DMs are small, to explore
 the hypothesis that the correlation between $z\sim 0.4$ galaxies and
 high-DM FRBs is due to propagation effects, rather than large host DMs.
 The host DM distribution (not shown) is sharply peaked at $\Dh \sim 70$ \pcc.}
\label{fig:dm_pdf}
\end{figure}

First, we model the distribution of FRBs in redshift and DM.
We assume that the FRB redshift distribution is
\be
\frac{dn_f^{2d}}{dz} \propto z^2 e^{-\gamma z}  \label{eq:dnfdz_fid}
\ee
and that the host DM distribution is lognormal, and independent 
of redshift:
\be
p(\Dh) \propto \frac{1}{\Dh} \exp\left( -\frac{(\log\Dh - \mu_{\log})^2}{2\sigma_{\log}^2} \right) \, .  \label{eq:pdh_fid}
\ee
In Eqs.~(\ref{eq:dnfdz_fid}),~(\ref{eq:pdh_fid}), we choose parameters
\be
\gamma = 6.7
 \hspace{1cm}
\mu_{\log} = 4.2
 \hspace{1cm}
\sigma_{\log} = 5 \, .  \label{eq:params_fid}
\ee
The total DM is $\DM = \Di(z) + \Dh$.
These parameters have been chosen so that
the median FRB redshift is 0.4,
the median host DM is 67 \pcc, and the distribution of total 
DMs is similar to the observed DM distribution 
in Figure~\ref{fig:dm_pdf}.

We will also need a fiducial model for $P_{ge}(k)$, the 3D galaxy-electron
power spectrum at comoving wavenumber $k$.
For reasons that we will explain shortly, we will need to
know the one-halo contribution in the limit $k\rightarrow 0$,
which is \citep{Rafiei-Ravandi:2020aa}
\be
\lim_{k \rightarrow 0} P_{ge}^{1h}(k,z) = \frac{\langle N_e^{\rm ion} \rangle}{n_{e,0}} \, ,\label{eq:pge_1h_k0}
\ee
where $\langle N_e^{\rm ion} \rangle$ is the average (over survey galaxies)
number of electrons in the halo containing a galaxy, and $n_{e,0}$
is the comoving electron number density.
To compute $\langle N_e^{\rm ion} \rangle$,
we assume that survey galaxies
are contained in dark matter halos whose mass $M_h$ is
lognormal-distributed, with parameters:
\be
\big\langle \lambda \big\rangle = 13.4
 \hspace{1cm}
\sigma\big(\lambda\big) = 0.35 \, ,
\ee
where $\lambda \equiv \log_{10}(M_h/M_\odot)$.
This distribution is a rough fit to the halo
mass distribution shown in Figure~3 
of~\cite{Collaboration:2021tt} for SDSS-LOWZ,
a well-characterized $z \sim 0.3$ galaxy survey
similar to the ones considered here.
We assume that these large halos have baryon-to-matter ratio
equal to the cosmic average $(\Omega_b/\Omega_m)$, with 
ionization fraction $f_b = 0.75$.

Finally, we model the CHIME/FRB selection function $S(\DM)$ in DM.
This has been measured via Monte Carlo analysis of simulated events,
and the result is shown in Figure 14 of the CHIME/FRB Catalog 1
paper \citep{Collaboration:2021wz}.
Here, we will use the following rough visual fit:
\be
\log S(\DM) = 0.1 - 0.14 \left[ \log\left( \frac{\DM}{1000} \right) \right]^2 .  \label{eq:dm_selection}
\ee
The selection function $S(\DM)$ is, up to normalization, the
probability that a random FRB with a given DM is detected by CHIME/FRB.
As an aside,
CHIME/FRB has a selection bias against detecting high-DM
FRBs due to frequency channel smearing
and a bias against detecting low-DM FRBs
due to the details of the high-pass filtering
used in radio frequency interference removal.
(Scattering biases will be discussed later in this section.)
This combination of biases results in the selection function~(Eq.~\ref{eq:dm_selection})
with a local maximum at $\DM \sim 1000$ \pcc.

With the fiducial model in the previous few paragraphs,
we now proceed to calculate contributions to 
\clfg\ from propagation effects.

The first propagation effect we will consider is ``DM-completeness'',
described schematically as follows.
Consider a foreground population of galaxies, and a background
(i.e.\ higher-redshift) population of FRBs.
The galaxies are spatially correlated with ionized electrons,
which increase DMs of the FRBs, by adding dispersion
along the line of sight.
This can either increase or decrease the apparent number density
of FRBs, depending on whether $dS/d(\DM)$ is positive or negative.
This combination of effects produces a correlation between number
densities of FRBs and galaxies, i.e.~a contribution to 
\clfg\ that can be positive or negative.

In~\citet{Rafiei-Ravandi:2020aa}, the contribution to 
\clfg\ from DM-completeness is calculated:
\be
C_\ell^{fg} = \frac{1}{n_g^{2d}} \int dz \, \frac{H(z)}{\chi(z)^2} \frac{dn_g^{2d}}{dz} W_f(z) \Pgelz  \, ,\label{eq:clfg_prop}
\ee
where the DM-completeness weight function $W_f$ for DM bin
$[\DM_{\rm min}, \DM_{\rm max}]$ is
\ba
W_f(z) &=&\frac{n_{e,0}}{n_f^{2d}} \, \frac{1+z}{H(z)} \int_z^\infty dz' \nn \\
&& \times \int_{\DM_{\rm min}}^{\DM_{\rm max}} d(\DM)
 \frac{d^2n^{2d}_f}{dz'\,d(\DM)} \frac{d\log S}{d(\DM)}  \label{eq:Wf_c} 
\ea
and $\chi(z)$ is comoving distance to redshift $z$.
We convert this expression
for \clfg\ to an expression for our parameter $\alpha$
as follows:
\begin{align}
\alpha &= \lim_{\ell\rightarrow 0} C_\ell^{fg(1h)} \nn \\
  &= \frac{1}{n_g^{2d}} \int dz \, \frac{H(z)}{\chi(z)^2} \frac{dn_g^{2d}}{dz} W_f(z) \frac{\langle N_e^{\rm ion}(z) \rangle}{n_{e,0}}  \, , \label{eq:alphaL_W}
\end{align}
where we have used Eq.~(\ref{eq:alpha_lim0}) in the first line
and Eqs.~(\ref{eq:pge_1h_k0}),~(\ref{eq:clfg_prop})
in the second line.

\begin{figure}
\centerline{\includegraphics[width=8.5cm]{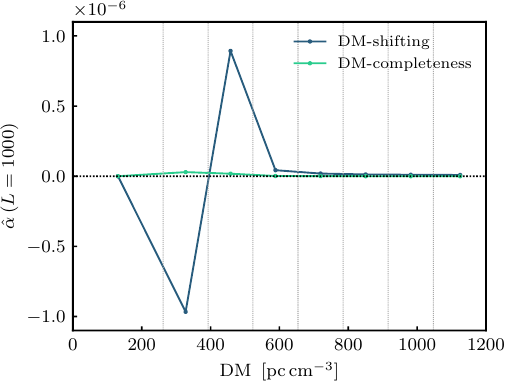}}
\caption{Predicted contribution to the FRB-galaxy correlation $\alpha_L$ 
(Eq.~\ref{eq:alpha-hat}) from propagation effects,
in the fiducial model from~\S\ref{ssec:propagation_effects}.
The DM binning is the same as Figure~\ref{fig:alpha_d}.
Comparing to the error bars in Figure~\ref{fig:alpha_d}, 
the DM-shifting contribution is $\sim 0.5\sigma$ in the second and
third DM bins ($262 < \DM < 393$ and $393 < \DM < 523$ \pcc)
and $\lesssim 0.1\sigma$ in the other bins.
The DM-completeness contribution is very small.}
\label{fig:propagation}
\end{figure}

The second propagation effect we will consider is ``DM-shifting'',
which arises for an FRB catalog that has been binned in DM,
as in Figure~\ref{fig:alpha_d}.
Even in the absence of an instrumental selection function,
DM fluctuations along the line of sight can shift FRBs
across DM bin boundaries, either increasing or decreasing
the observed number density of FRBs in a given bin.
This effect is distinct from the DM-completeness effect
described above, and also produces a contribution to 
\clfg\ that can be positive or negative.
Using results from~\cite{Rafiei-Ravandi:2020aa}, the
DM-shifting bias to $\alpha_L$ is given by the
previous expression~(\ref{eq:alphaL_W}), but with the
following expression for the DM-shifting weight function:
\be
W_f(z) = -\frac{n_{e,0}}{n_f^{2d}} \, \frac{1+z}{H(z)} \int_z^\infty dz' 
\left[ \frac{d^2n_f^{2d}}{dz' \, d(\DM)} \right]_{\DM_{\rm min}}^{\DM_{\rm max}} \, .  \label{eq:Wf_ds}
\ee
In Figure~\ref{fig:propagation}, we show $\alpha_L$-biases
from the DM-completeness and DM-shifting propagation effects
in our fiducial model, computed using Eqs.~(\ref{eq:Wf_c})--(\ref{eq:Wf_ds}).
For simplicity, we have approximated the precise $z$-dependence
of the redshift-binned galaxy surveys in Figure~\ref{fig:alpha_d}
by assuming $dn_g^{2d}/dz=\mbox{const}$ for $0.3 \le z \le 0.4$.
(The results are not very sensitive to the galaxy redshift distribution.)

Comparing to the FRB-galaxy correlation shown previously in Figure~\ref{fig:alpha_d},
we see that the total bias is $\sim 0.5\sigma$ in the second DM bin
($262 < \DM < 393$ \pcc), and $\lesssim 0.1\sigma$ in the other bins.
These biases are too small, and have the wrong DM dependence, to
explain the FRB-galaxy correlation shown previously in Figure~\ref{fig:alpha_d}.

So far, we have only considered propagation effects involving dispersion.
The next propagation effect we might want to consider is scattering
completeness, described intuitively as follows.
Consider a foreground population of galaxies
and a background population of FRBs.
The galaxies are correlated
with free electrons, which scatter-broaden FRBs and change their
observed number density.
Since scatter-broadening always decreases the probability that an
FRB is detected, this effect always produces negative \clfg.\footnote{Formally, the selection function for scattering is a decreasing
function of scattering width. This can be seen directly in
Figure~15 of \citep{Collaboration:2021wz}.}
Therefore, scattering completeness cannot be
responsible for the observed FRB-galaxy correlation,
which is positive (as expected for clustering).

A final category of propagation effects is strong lensing
(either plasma lensing or gravitational lensing) by foreground
galaxies.
Although strong lenses are rare, they can produce
large magnification, increasing the detection rate
of background FRBs by a large factor if the FRB luminosity
function is sufficiently steep.
A complete analysis of strong lensing in CHIME/FRB would be
a substantial undertaking, and we defer it to a future paper.

\section{Summary and conclusions}
\label{sec:discussion}

In this paper, we find a cross-correlation
between CHIME FRBs and galaxies at redshifts $0.3 \lesssim z \lesssim 0.5$.
The correlation is statistically significant in three galaxy surveys:
WISE$\times$SCOS, DESI-BGS, and DESI-LRG.
The statistical significance of the detection in each survey is 
$p \sim 2.7 \times 10^{-5}$, $3.1 \times 10^{-4}$, and $4.1 \times 10^{-4}$,
respectively.
These \pvalues account for look-elsewhere effects in
both angular scale $L$ and redshift range.

The FRB-galaxy correlation is detected on angular scales ($\ell \sim 1000$)
in the one-halo regime.
In this regime, the amplitude of the correlation is proportional
to the mean ``link count'' $\eta$ of the FRB population, i.e.~mean number
of galaxies in the same halo as an FRB.
Cross-correlating CHIME FRBs with $0.3 \lesssim z \lesssim 0.5$ galaxies,
we find $\eta$ of order unity.

This measurement of $\eta$ cannot be directly translated to the
probability $p$ that an FRB is in the given redshift range.
We can write $\eta = p \tilde \eta$, where $\tilde\eta$ is the
mean link count of FRBs in the redshift range.
Formally, we measure $(p\tilde\eta)$ but not the individual factors
$p,\tilde\eta$.
However, in the bright galaxy surveys considered here,
dark matter halos rarely contain more than a few
catalog galaxies.
We conclude that $\tilde\eta$ must be of order unity, implying that 
$p$ is also of order unity. 
That is, an order-one fraction of CHIME FRBs are in redshift
range $0.3 \lesssim z \lesssim 0.5$.

We have phrased this conclusion as a qualitative statement 
(``order-one fraction''), since it is difficult to assign a
quantitative upper bound to $\tilde\eta$.
This issue is a limitation of measuring FRB-galaxy correlations
in the one-halo regime, where the FRB redshift distribution
always appears multiplied by a linking factor $\tilde\eta$.
Future CHIME/FRB catalogs should contain enough FRBs to detect the FRB-galaxy
correlation on two-halo scales ($\ell \sim 100$)~\citep{Rafiei-Ravandi:2020aa},
which will help break this degeneracy.

We find statistical evidence for a population of FRBs with
large host DMs, on the order of $\Dh \sim 400$ \pcc.
More precisely, we detect a nonzero correlation between FRBs
with DM $\ge 785$ \pcc\ (after subtracting the YWM16 estimate
of the Milky Way DM) and galaxies at $z \sim 0.4$, where
the IGM contribution to the DM is $\Di(z) \sim 360$ \pcc.

This may appear to be in tension with direct host galaxy associations.
At the time of this writing, 14 FRBs have been localized to host galaxies,
all of which have $\Dh \lesssim 200$ \pcc.
However, FRBs with DM $\ge 785$ \pcc\ are currently uncommon, and our FRB-galaxy
correlation result must be interpreted carefully.
It implies that an order-one fraction of high-DM FRBs
are at redshift $z \sim 0.4$ in CHIME/FRB, but it does not imply that an order-one fraction
of FRBs at redshift $z \sim 0.4$ have high DM.
These statements are actually consistent with the direct associations.
Since there is currently only one direct association with YMW16-subtracted DM $\ge 785$ \pcc,
one cannot currently rule out the possibility that an order-one
fraction of high-DM FRBs are at $z \sim 0.4$.

The number of direct host associations is rapidly growing, and
we predict that direct associations will soon find high-DM
FRBs with $z \sim 0.4$.
However, we note that most direct associations to date have
been discovered by ASKAP at lower DM (on average) than the CHIME/FRB
sample.

We briefly explore mechanisms for producing host DMs $\gtrsim 400$ \pcc,
and show that contributions from gas in large halos provide a plausible
mechanism.
Quantitatively, we find that for FRBs near the centers ($r \lesssim 100$ kpc)
of large ($M \sim 10^{14} M_\odot$) halos the host halo DM can be $\gtrsim 400$
\pcc\ (Figure~\ref{fig:Dhh}), at least in one widely used ICM model \citep{Prochaska:2019ab}.
FRBs in such halos will strongly correlate with galaxies, since bright
survey galaxies are often found in large halos.
We show that line-of-sight propagation effects are unlikely to be a significant
source of bias (\S\ref{ssec:propagation_effects}).

Future measurements of FRB-galaxy cross-correlations will
have higher SNR, and the results presented here
could be extended in several ways. One could bin
simultaneously in galaxy redshift and FRB DM, 
to explore the FRB-galaxy correlation strength as a function
of two variables $(z,\mbox{DM})$.
Cross-correlations can constrain the high-$z$ tail
of the FRB redshift distribution, where direct associations 
are difficult since individual galaxies are usually 
faint \citep{Eftekhari:2017aa}.
Very high-$z$ FRBs, if present, can be used to constrain
cosmic reionization history
\citep{Caleb:2019tg,Linder:2020ul,Zhang:2021aa}.
Finally, line-of-sight propagation effects
will eventually be detectable in \clfg, and will be
an interesting probe of the distribution of electrons in the universe.

This paper is based on FRBs from CHIME/FRB Catalog 1,
which contains 489 unique sources and approximate angular sky positions.
Future CHIME/FRB catalogs will include more FRB sources,
many of which will have improved angular resolution
through use of baseband data \citep{Michilli:2021tm}.
The FRB-galaxy correlation presented here should have
much higher statistical significance in future CHIME/FRB
catalogs and will be exciting to explore.

\acknowledgements
We thank the Dominion Radio Astrophysical Observatory, operated by the National Research Council Canada, for gracious
hospitality and expertise. CHIME is funded by a grant from the Canada Foundation for Innovation (CFI) 2012 Leading Edge
Fund (Project 31170) and by contributions from the provinces of British Columbia, Qu\'ebec and Ontario. The CHIME/FRB
Project is funded by a grant from the CFI 2015 Innovation Fund (Project 33213), and by contributions from the provinces of
British Columbia, Qu\'ebec, and by the Dunlap Institute for Astronomy and Astrophysics at the University of Toronto.
The Dunlap Institute is funded through an endowment established by the David Dunlap family and the University of Toronto.
Additional support was provided by the Canadian Institute for Advanced Research (CIFAR), McGill University and the McGill
Space Institute via the Trottier Family Foundation, and the University of British Columbia. Research at Perimeter Institute is
supported by the Government of Canada through Industry Canada and by the Province of Ontario through the Ministry of
Research \& Innovation.
K.M.S. was supported by an NSERC Discovery Grant and a CIFAR fellowship.
\acks
This research was enabled in part by support provided by WestGrid (www.westgrid.ca)
and Compute Canada (www.computecanada.ca).
The Photometric Redshifts for the Legacy Surveys (PRLS) catalog used in this paper was
produced thanks to funding from the U.S. Department of Energy Office of Science, Office
of High Energy Physics via grant DE-SC0007914.

\bibliographystyle{aasjournal}
\bibliography{ch_frbx_c1.bib}

\begin{thebibliography}{}
\expandafter\ifx\csname natexlab\endcsname\relax\def\natexlab#1{#1}\fi
\providecommand{\url}[1]{\href{#1}{#1}}
\providecommand{\dodoi}[1]{doi:~\href{http://doi.org/#1}{\nolinkurl{#1}}}
\providecommand{\doeprint}[1]{\href{http://ascl.net/#1}{\nolinkurl{http://ascl.net/#1}}}
\providecommand{\doarXiv}[1]{\href{https://arxiv.org/abs/#1}{\nolinkurl{https://arxiv.org/abs/#1}}}

\bibitem[{Aghanim {et~al.}(2020)Aghanim, Akrami, Ashdown, Aumont, Baccigalupi,
  Ballardini, Banday, Barreiro, Bartolo, \& et~al.}]{Planck_2018}
Aghanim, N., Akrami, Y., Ashdown, M., {et~al.} 2020, A\&A, 641, A6,
  \dodoi{10.1051/0004-6361/201833910}

\bibitem[{Alonso(2021)}]{Alonso:2021aa}
Alonso, D. 2021, Physical Review D, 103, 123544,
  \dodoi{10.1103/PhysRevD.103.123544}

\bibitem[{Alonso {et~al.}(2015)Alonso, Salvador, S{\'a}nchez, Bilicki,
  Garc{\'\i}a-Bellido, \& S{\'a}nchez}]{Alonso:2015aa}
Alonso, D., Salvador, A.~I., S{\'a}nchez, F.~J., {et~al.} 2015, Monthly Notices
  of the Royal Astronomical Society, 449, 670, \dodoi{10.1093/mnras/stv309}

\bibitem[{Balaguera-Antol{\'\i}nez {et~al.}(2018)Balaguera-Antol{\'\i}nez,
  Bilicki, Branchini, \& Postiglione}]{Balaguera-Antolinez:2018aa}
Balaguera-Antol{\'\i}nez, A., Bilicki, M., Branchini, E., \& Postiglione, A.
  2018, Monthly Notices of the Royal Astronomical Society, 476, 1050,
  \dodoi{10.1093/mnras/sty262}

\bibitem[{Bannister {et~al.}(2019)Bannister, Deller, Phillips, Macquart,
  Prochaska, Tejos, Ryder, Sadler, Shannon, Simha, Day, McQuinn, North-Hickey,
  Bhandari, Arcus, Bennert, Burchett, Bouwhuis, Dodson, Ekers, Farah, Flynn,
  James, Kerr, Lenc, Mahony, O'Meara, Os{\l}owski, Qiu, Treu, U, Bateman, Bock,
  Bolton, Brown, Bunton, Chippendale, Cooray, Cornwell, Gupta, Hayman,
  Kesteven, Koribalski, MacLeod, McClure-Griffiths, Neuhold, Norris, Pilawa,
  Qiao, Reynolds, Roxby, Shimwell, Voronkov, \& Wilson}]{Bannister:2019aa}
Bannister, K.~W., Deller, A.~T., Phillips, C., {et~al.} 2019, Science, 365,
  565, \dodoi{10.1126/science.aaw5903}

\bibitem[{Bassa {et~al.}(2017)Bassa, Tendulkar, Adams, Maddox, Bogdanov, Bower,
  Burke-Spolaor, Butler, Chatterjee, Cordes, Hessels, Kaspi, Law, Marcote,
  Paragi, Ransom, Scholz, Spitler, \& Langevelde}]{Bassa:2017uf}
Bassa, C.~G., Tendulkar, S.~P., Adams, E. A.~K., {et~al.} 2017, The
  Astrophysical Journal, 843, L8, \dodoi{10.3847/2041-8213/aa7a0c}

\bibitem[{Bhandari {et~al.}(2020{\natexlab{a}})Bhandari, Sadler, Prochaska,
  Simha, Ryder, Marnoch, Bannister, Macquart, Flynn, Shannon, Tejos,
  Corro-Guerra, Day, Deller, Ekers, Lopez, Mahony, Nu{\~n}ez, \&
  Phillips}]{Bhandari:2020aa}
Bhandari, S., Sadler, E.~M., Prochaska, J.~X., {et~al.} 2020{\natexlab{a}}, The
  Astrophysical Journal, 895, L37, \dodoi{10.3847/2041-8213/ab672e}

\bibitem[{Bhandari {et~al.}(2020{\natexlab{b}})Bhandari, Bannister, Lenc, Cho,
  Ekers, Day, Deller, Flynn, James, Macquart, Mahony, Marnoch, Moss, Phillips,
  Prochaska, Qiu, Ryder, Shannon, Tejos, \& Wong}]{Bhandari:2020uj}
Bhandari, S., Bannister, K.~W., Lenc, E., {et~al.} 2020{\natexlab{b}}, The
  Astrophysical Journal, 901, L20, \dodoi{10.3847/2041-8213/abb462}

\bibitem[{Bhardwaj {et~al.}(2021)Bhardwaj, Gaensler, Kaspi, Landecker,
  Mckinven, Michilli, Pleunis, Tendulkar, Andersen, Boyle, Cassanelli, Chawla,
  Cook, Dobbs, Fonseca, Kaczmarek, Leung, Masui, Mnchmeyer, Ng, Rafiei-Ravandi,
  Scholz, Shin, Smith, Stairs, \& Zwaniga}]{Bhardwaj:2021aa}
Bhardwaj, M., Gaensler, B.~M., Kaspi, V.~M., {et~al.} 2021, The Astrophysical
  Journal, 910, L18, \dodoi{10.3847/2041-8213/abeaa6}

\bibitem[{Bilicki {et~al.}(2013)Bilicki, Jarrett, Peacock, Cluver, \&
  Steward}]{Bilicki:2013aa}
Bilicki, M., Jarrett, T.~H., Peacock, J.~A., Cluver, M.~E., \& Steward, L.
  2013, The Astrophysical Journal Supplement Series, 210, 9,
  \dodoi{10.1088/0067-0049/210/1/9}

\bibitem[{Bilicki {et~al.}(2016)Bilicki, Peacock, Jarrett, Cluver, Maddox,
  Brown, Taylor, Hambly, Solarz, Holwerda, Baldry, Loveday, Moffett, Hopkins,
  Driver, Alpaslan, \& Bland-Hawthorn}]{Bilicki:2016aa}
Bilicki, M., Peacock, J.~A., Jarrett, T.~H., {et~al.} 2016, The Astrophysical
  Journal Supplement Series, 225, 5, \dodoi{10.3847/0067-0049/225/1/5}

\bibitem[{Bochenek {et~al.}(2020)Bochenek, Ravi, Belov, Hallinan, Kocz,
  Kulkarni, \& McKenna}]{Bochenek_2020}
Bochenek, C.~D., Ravi, V., Belov, K.~V., {et~al.} 2020, Nature, 587, 59,
  \dodoi{10.1038/s41586-020-2872-x}

\bibitem[{Caleb {et~al.}(2019)Caleb, Flynn, \& Stappers}]{Caleb:2019tg}
Caleb, M., Flynn, C., \& Stappers, B.~W. 2019, Monthly Notices of the Royal
  Astronomical Society, 485, 2281, \dodoi{10.1093/mnras/stz571}

\bibitem[{Chatterjee {et~al.}(2017)Chatterjee, Law, Wharton, Burke-Spolaor,
  Hessels, Bower, Cordes, Tendulkar, Bassa, Demorest, Butler, Seymour, Scholz,
  Abruzzo, Bogdanov, Kaspi, Keimpema, Lazio, Marcote, McLaughlin, Paragi,
  Ransom, Rupen, Spitler, \& van Langevelde}]{Chatterjee:2017aa}
Chatterjee, S., Law, C.~J., Wharton, R.~S., {et~al.} 2017, Nature, 541, 58,
  \dodoi{10.1038/nature20797}

\bibitem[{{CHIME/FRB Collaboration}(2018)}]{Collaboration:2018aa}
{CHIME/FRB Collaboration}. 2018, The Astrophysical Journal, 863, 48,
  \dodoi{10.3847/1538-4357/aad188}

\bibitem[{{CHIME/FRB Collaboration}(2019)}]{Collaboration:2019tw}
---. 2019, The Astrophysical Journal, 885, L24,
  \dodoi{10.3847/2041-8213/ab4a80}

\bibitem[{{CHIME/FRB Collaboration}(2020{\natexlab{a}})}]{chimfrb:2020vo}
---. 2020{\natexlab{a}}, Nature, 587, 54, \dodoi{10.1038/s41586-020-2863-y}

\bibitem[{{CHIME/FRB Collaboration}(2020{\natexlab{b}})}]{Collaboration:2020vl}
---. 2020{\natexlab{b}}, Nature, 582, 351, \dodoi{10.1038/s41586-020-2398-2}

\bibitem[{{CHIME/FRB Collaboration}(2021)}]{Collaboration:2021wz}
---. 2021, Submitted to ApJS

\bibitem[{Chittidi {et~al.}(2020)Chittidi, Simha, Mannings, Prochaska,
  Rafelski, Neeleman, Macquart, Tejos, Jorgenson, Ryder, Day, Marnoch,
  Bhandari, Deller, Qiu, Bannister, Shannon, \& Heintz}]{Chittidi:2020aa}
Chittidi, J.~S., Simha, S., Mannings, A., {et~al.} 2020, arXiv e-prints,
  arXiv:2005.13158.
\newblock \url{https://ui.adsabs.harvard.edu/abs/2020arXiv200513158C}

\bibitem[{Cooray \& Sheth(2002)}]{COORAY_2002}
Cooray, A., \& Sheth, R. 2002, Physics Reports, 372, 1,
  \dodoi{10.1016/s0370-1573(02)00276-4}

\bibitem[{Cordes \& Chatterjee(2019)}]{Cordes:2019wl}
Cordes, J.~M., \& Chatterjee, S. 2019, Annual Review of Astronomy and
  Astrophysics, 57, 417, \dodoi{10.1146/annurev-astro-091918-104501}

\bibitem[{Cordes \& Lazio(2002)}]{Cordes:2002tt}
Cordes, J.~M., \& Lazio, T. J.~W. 2002, arXiv e-prints, astro.
\newblock \url{https://ui.adsabs.harvard.edu/abs/2002astro.ph..7156C}

\bibitem[{Dey {et~al.}(2019)Dey, Schlegel, Lang, Blum, Burleigh, Fan, Findlay,
  Finkbeiner, Herrera, Juneau, Landriau, Levi, McGreer, Meisner, Myers,
  Moustakas, Nugent, Patej, Schlafly, Walker, Valdes, Weaver, Y{\`{e}}che, Zou,
  Zhou, Abareshi, Abbott, Abolfathi, Aguilera, Alam, Allen, Alvarez, Annis,
  Ansarinejad, Aubert, Beechert, Bell, BenZvi, Beutler, Bielby, Bolton,
  Brice{\~{n}}o, Buckley-Geer, Butler, Calamida, Carlberg, Carter, Casas,
  Castander, Choi, Comparat, Cukanovaite, Delubac, DeVries, Dey, Dhungana,
  Dickinson, Ding, Donaldson, Duan, Duckworth, Eftekharzadeh, Eisenstein,
  Etourneau, Fagrelius, Farihi, Fitzpatrick, Font-Ribera, Fulmer, G{\"a}nsicke,
  Gaztanaga, George, Gerdes, Gontcho, Gorgoni, Green, Guy, Harmer, Hernandez,
  Honscheid, Huang, James, Jannuzi, Jiang, Joyce, Karcher, Karkar, Kehoe,
  Jean-Paul, Kueter-Young, Lan, Lauer, Guillou, Suu, Lee, Lesser, Levasseur,
  Li, Mann, Marshall, Mart{\'{\i}}nez-V{\'{a}}zquez, Martini, du~Mas~des
  Bourboux, McManus, Meier, M{\'{e}}nard, Metcalfe,
  Mu{\~{n}}oz-Guti{\'{e}}rrez, Najita, Napier, Narayan, Newman, Nie, Nord,
  Norman, Olsen, Paat, Palanque-Delabrouille, Peng, Poppett, Poremba, Prakash,
  Rabinowitz, Raichoor, Rezaie, Robertson, Roe, Ross, Ross, Rudnick, Safonova,
  Saha, S{\'{a}}nchez, Savary, Schweiker, Scott, Seo, Shan, Silva, Slepian,
  Soto, Sprayberry, Staten, Stillman, Stupak, Summers, Tie, Tirado,
  Vargas-Maga{\~{n}}a, Vivas, Wechsler, Williams, Yang, Yang, Yapici, Zaritsky,
  Zenteno, Zhang, Zhang, Zhou, \& Zhou}]{Dey:2019aa}
Dey, A., Schlegel, D.~J., Lang, D., {et~al.} 2019, The Astronomical Journal,
  157, 168, \dodoi{10.3847/1538-3881/ab089d}

\bibitem[{Eftekhari \& Berger(2017)}]{Eftekhari:2017aa}
Eftekhari, T., \& Berger, E. 2017, The Astrophysical Journal, 849, 162,
  \dodoi{10.3847/1538-4357/aa90b9}

\bibitem[{Gorski {et~al.}(2005)Gorski, Hivon, Banday, Wandelt, Hansen,
  Reinecke, \& Bartelmann}]{Gorski:2005aa}
Gorski, K.~M., Hivon, E., Banday, A.~J., {et~al.} 2005, The Astrophysical
  Journal, 622, 759, \dodoi{10.1086/427976}

\bibitem[{Heintz {et~al.}(2020)Heintz, Prochaska, Simha, Platts, Fong, Tejos,
  Ryder, Aggerwal, Bhandari, Day, Deller, Kilpatrick, Law, Macquart, Mannings,
  Marnoch, Sadler, \& Shannon}]{Heintz:2020aa}
Heintz, K.~E., Prochaska, J.~X., Simha, S., {et~al.} 2020, The Astrophysical
  Journal, 903, 152, \dodoi{10.3847/1538-4357/abb6fb}

\bibitem[{Hodges(1958)}]{Hodges:1958aa}
Hodges, J.~L. 1958, Ark. Mat., 3, 469, \dodoi{10.1007/BF02589501}

\bibitem[{James {et~al.}(2021{\natexlab{a}})James, Prochaska, Macquart,
  North-Hickey, Bannister, \& Dunning}]{James:2021aa}
James, C.~W., Prochaska, J.~X., Macquart, J.~P., {et~al.} 2021{\natexlab{a}},
  arXiv e-prints, arXiv:2101.07998.
\newblock \url{https://ui.adsabs.harvard.edu/abs/2021arXiv210107998J}

\bibitem[{James {et~al.}(2021{\natexlab{b}})James, Prochaska, Macquart,
  North-Hickey, Bannister, \& Dunning}]{James:2021ab}
---. 2021{\natexlab{b}}, arXiv e-prints, arXiv:2101.08005.
\newblock \url{https://ui.adsabs.harvard.edu/abs/2021arXiv210108005J}

\bibitem[{Josephy {et~al.}(2021)Josephy, Chawla, Curtin, Kaspi, Bhardwaj,
  Boyle, Brar, Cassanelli, Fonseca, Gaensler, Leung, Lin, Masui, McKinven,
  Mena-Parra, Michilli, Ng, Pleunis, Rafiei-Ravandi, Rahman, Sanghavi, Scholz,
  Smith, Stairs, \& Tendulkar}]{Josephy:2021ts}
Josephy, A., Chawla, P., Curtin, A.~P., {et~al.} 2021, Submitted to ApJ

\bibitem[{Keating \& Pen(2020)}]{Keating:2020aa}
Keating, L.~C., \& Pen, U.-L. 2020, Monthly Notices of the Royal Astronomical
  Society: Letters, 496, L106, \dodoi{10.1093/mnrasl/slaa095}

\bibitem[{Kokubo {et~al.}(2017)Kokubo, Mitsuda, Sugai, Ozaki, Minowa, Hattori,
  Hayano, Matsubayashi, Shimono, Sako, \& Doi}]{Kokubo:2017uy}
Kokubo, M., Mitsuda, K., Sugai, H., {et~al.} 2017, The Astrophysical Journal,
  844, 95, \dodoi{10.3847/1538-4357/aa7b2d}

\bibitem[{Krakowski {et~al.}(2016)Krakowski, Ma{\l}ek, Bilicki, Pollo, Kurcz,
  \& Krupa}]{Krakowski:2016aa}
Krakowski, T., Ma{\l}ek, K., Bilicki, M., {et~al.} 2016, A\&A, 596.
\newblock \url{https://doi.org/10.1051/0004-6361/201629165}

\bibitem[{Law {et~al.}(2020)Law, Butler, Prochaska, Zackay, Burke-Spolaor,
  Mannings, Tejos, Josephy, Andersen, Chawla, Heintz, Aggarwal, Bower,
  Demorest, Kilpatrick, Lazio, Linford, Mckinven, Tendulkar, \&
  Simha}]{Law:2020aa}
Law, C.~J., Butler, B.~J., Prochaska, J.~X., {et~al.} 2020, The Astrophysical
  Journal, 899, 161, \dodoi{10.3847/1538-4357/aba4ac}

\bibitem[{Li {et~al.}(2019)Li, Yalinewich, \& Breysse}]{Li:2019fsg}
Li, D., Yalinewich, A., \& Breysse, P.~C. 2019, arXiv e-prints,
  arXiv:1902.10120

\bibitem[{Linder(2020)}]{Linder:2020ul}
Linder, E.~V. 2020, Physical Review D, 101, 103019,
  \dodoi{10.1103/PhysRevD.101.103019}

\bibitem[{Macquart {et~al.}(2020)Macquart, Prochaska, McQuinn, Bannister,
  Bhandari, Day, Deller, Ekers, James, Marnoch, Os{\l}owski, Phillips, Ryder,
  Scott, Shannon, \& Tejos}]{Macquart:2020aa}
Macquart, J.~P., Prochaska, J.~X., McQuinn, M., {et~al.} 2020, Nature, 581,
  391, \dodoi{10.1038/s41586-020-2300-2}

\bibitem[{Madhavacheril {et~al.}(2019)Madhavacheril, Battaglia, Smith, \&
  Sievers}]{Madhavacheril:2019vm}
Madhavacheril, M.~S., Battaglia, N., Smith, K.~M., \& Sievers, J.~L. 2019,
  Physical Review D, 100, 103532, \dodoi{10.1103/PhysRevD.100.103532}

\bibitem[{Mannings {et~al.}(2021)Mannings, Fong, Simha, Prochaska, Rafelski,
  Kilpatrick, Tejos, Heintz, Bannister, Bhandari, Day, Deller, Ryder, Shannon,
  \& Tendulkar}]{Mannings:2021ws}
Mannings, A.~G., Fong, W.-f., Simha, S., {et~al.} 2021, The Astrophysical
  Journal, 917, 75, \dodoi{10.3847/1538-4357/abff56}

\bibitem[{Marcote {et~al.}(2020)Marcote, Nimmo, Hessels, Tendulkar, Bassa,
  Paragi, Keimpema, Bhardwaj, Karuppusamy, Kaspi, Law, Michilli, Aggarwal,
  Andersen, Archibald, Bandura, Bower, Boyle, Brar, Burke-Spolaor, Butler,
  Cassanelli, Chawla, Demorest, Dobbs, Fonseca, Giri, Good, Gourdji, Josephy,
  Kirichenko, Kirsten, Landecker, Lang, Lazio, Li, Lin, Linford, Masui,
  Mena-Parra, Naidu, Ng, Patel, Pen, Pleunis, Rafiei-Ravandi, Rahman, Renard,
  Scholz, Siegel, Smith, Stairs, Vanderlinde, \& Zwaniga}]{Marcote:2020ts}
Marcote, B., Nimmo, K., Hessels, J. W.~T., {et~al.} 2020, Nature, 577, 190,
  \dodoi{10.1038/s41586-019-1866-z}

\bibitem[{Masui \& Sigurdson(2015)}]{Masui:2015wu}
Masui, K.~W., \& Sigurdson, K. 2015, Physical Review Letters, 115, 121301,
  \dodoi{10.1103/PhysRevLett.115.121301}

\bibitem[{McQuinn(2014)}]{McQuinn:2014aa}
McQuinn, M. 2014, The Astrophysical Journal Letters, 780, L33,
  \dodoi{10.1088/2041-8205/780/2/l33}

\bibitem[{Michilli {et~al.}(2021)Michilli, Masui, Mckinven, Cubranic,
  Bruneault, Brar, Patel, Boyle, Stairs, Renard, Bandura, Berger, Breitman,
  Cassanelli, Dobbs, Kaspi, Leung, Mena-Parra, Pleunis, Russell, Scholz,
  Siegel, Tendulkar, \& Vanderlinde}]{Michilli:2021tm}
Michilli, D., Masui, K.~W., Mckinven, R., {et~al.} 2021, The Astrophysical
  Journal, 910, 147, \dodoi{10.3847/1538-4357/abe626}

\bibitem[{Navarro {et~al.}(1997)Navarro, Frenk, \& White}]{Navarro:1996gj}
Navarro, J.~F., Frenk, C.~S., \& White, S. D.~M. 1997, Astrophys. J., 490, 493,
  \dodoi{10.1086/304888}

\bibitem[{Petroff {et~al.}(2019)Petroff, Hessels, \& Lorimer}]{Petroff:2019vo}
Petroff, E., Hessels, J. W.~T., \& Lorimer, D.~R. 2019, Astronomy and
  Astrophysics Review, 27, 4, \dodoi{10.1007/s00159-019-0116-6}

\bibitem[{{Petroff} \& {Yaron}(2020)}]{2020TNSAN.160....1P}
{Petroff}, E., \& {Yaron}, O. 2020, Transient Name Server AstroNote, 160, 1

\bibitem[{Platts {et~al.}(2019)Platts, Weltman, Walters, Tendulkar, Gordin, \&
  Kandhai}]{Platts:2019ws}
Platts, E., Weltman, A., Walters, A., {et~al.} 2019, Physics Reports, 821, 1,
  \dodoi{10.1016/j.physrep.2019.06.003}

\bibitem[{Prochaska \& Zheng(2019)}]{Prochaska:2019ab}
Prochaska, J.~X., \& Zheng, Y. 2019, Monthly Notices of the Royal Astronomical
  Society, 485, 648, \dodoi{10.1093/mnras/stz261}

\bibitem[{Prochaska {et~al.}(2019)Prochaska, Macquart, McQuinn, Simha, Shannon,
  Day, Marnoch, Ryder, Deller, Bannister, Bhandari, Bordoloi, Bunton, Cho,
  Flynn, Mahony, Phillips, Qiu, \& Tejos}]{Prochaska:2019aa}
Prochaska, J.~X., Macquart, J.-P., McQuinn, M., {et~al.} 2019, Science, 366,
  231, \dodoi{10.1126/science.aay0073}

\bibitem[{Rafiei-Ravandi {et~al.}(2020)Rafiei-Ravandi, Smith, \&
  Masui}]{Rafiei-Ravandi:2020aa}
Rafiei-Ravandi, M., Smith, K.~M., \& Masui, K.~W. 2020, Physical Review D, 102,
  023528, \dodoi{10.1103/PhysRevD.102.023528}

\bibitem[{Raichoor {et~al.}(2020)Raichoor, Eisenstein, Karim, Newman,
  Moustakas, Brooks, Dawson, Dey, Duan, Eftekharzadeh, Gazta{\~n}aga, Kehoe,
  Landriau, Lang, Lee, Levi, Meisner, Myers, Palanque-Delabrouille, Poppett,
  Prada, Ross, Schlegel, Schubnell, Staten, Tarl{\'e}, Tojeiro, Y{\`e}che, \&
  Zhou}]{Raichoor:2020aa}
Raichoor, A., Eisenstein, D.~J., Karim, T., {et~al.} 2020, Research Notes of
  the American Astronomical Society, 4, 180, \dodoi{10.3847/2515-5172/abc078}

\bibitem[{Ravi {et~al.}(2019)Ravi, Catha, D'Addario, Djorgovski, Hallinan,
  Hobbs, Kocz, Kulkarni, Shi, Vedantham, Weinreb, \& Woody}]{Ravi:2019aa}
Ravi, V., Catha, M., D'Addario, L., {et~al.} 2019, Nature, 572, 352,
  \dodoi{10.1038/s41586-019-1389-7}

\bibitem[{Reischke {et~al.}(2021{\natexlab{a}})Reischke, Hagstotz, \&
  Lilow}]{Reischke:2021th}
Reischke, R., Hagstotz, S., \& Lilow, R. 2021{\natexlab{a}}, Physical Review D,
  103, 023517, \dodoi{10.1103/PhysRevD.103.023517}

\bibitem[{Reischke {et~al.}(2021{\natexlab{b}})Reischke, Hagstotz, \&
  Lilow}]{Reischke:2021vx}
---. 2021{\natexlab{b}}, arXiv e-prints, arXiv:2102.11554.
\newblock \url{https://ui.adsabs.harvard.edu/abs/2021arXiv210211554R}

\bibitem[{Ruiz-Macias {et~al.}(2020)Ruiz-Macias, Zarrouk, Cole, Norberg, Baugh,
  Brooks, Dey, Duan, Eftekharzadeh, Eisenstein, Forero-Romero, Gazta{\~n}aga,
  Hahn, Kehoe, Landriau, Lang, Levi, Lucey, Meisner, Moustakas, Myers,
  Palanque-Delabrouille, Poppett, Prada, Raichoor, Schlegel, Schubnell,
  Tarl{\'e}, Weinberg, Wilson, \& Y{\`e}che}]{Ruiz-Macias:2020aa}
Ruiz-Macias, O., Zarrouk, P., Cole, S., {et~al.} 2020, Research Notes of the
  American Astronomical Society, 4, 187, \dodoi{10.3847/2515-5172/abc25a}

\bibitem[{Schaan {et~al.}(2021)Schaan, Ferraro, Amodeo, Battaglia, Aiola,
  Austermann, Beall, Bean, Becker, Bond, Calabrese, Calafut, Choi, Denison,
  Devlin, Duff, Duivenvoorden, Dunkley, D{\"u}nner, Gallardo, Guan, Han, Hill,
  Hilton, Hilton, Hlo{\v z}ek, Hubmayr, Huffenberger, Hughes, Koopman,
  MacInnis, McMahon, Madhavacheril, Moodley, Mroczkowski, Naess, Nati,
  Newburgh, Niemack, Page, Partridge, Salatino, Sehgal, Schillaci, Sif{\'o}n,
  Smith, Spergel, Staggs, Storer, Trac, Ullom, Van~Lanen, Vale, van Engelen,
  Maga{\~n}a, Vavagiakis, Wollack, \& Xu}]{Collaboration:2021tt}
Schaan, E., Ferraro, S., Amodeo, S., {et~al.} 2021, Physical Review D, 103,
  063513, \dodoi{10.1103/PhysRevD.103.063513}

\bibitem[{Scholz \& Stephens(1987)}]{Scholz:1987aa}
Scholz, F.~W., \& Stephens, M.~A. 1987, Journal of the American Statistical
  Association, 82, 918, \dodoi{10.1080/01621459.1987.10478517}

\bibitem[{Shirasaki {et~al.}(2017)Shirasaki, Kashiyama, \&
  Yoshida}]{Shirasaki:2017vd}
Shirasaki, M., Kashiyama, K., \& Yoshida, N. 2017, Physical Review D, 95,
  083012, \dodoi{10.1103/PhysRevD.95.083012}

\bibitem[{Simha {et~al.}(2020)Simha, Burchett, Prochaska, Chittidi, Elek,
  Tejos, Jorgenson, Bannister, Bhandari, Day, Deller, Forbes, Macquart, Ryder,
  \& Shannon}]{Simha:2020ug}
Simha, S., Burchett, J.~N., Prochaska, J.~X., {et~al.} 2020, The Astrophysical
  Journal, 901, 134, \dodoi{10.3847/1538-4357/abafc3}

\bibitem[{Spanakis-Misirlis(2021)}]{Spanakis-Misirlis:2021ta}
Spanakis-Misirlis, A. 2021, Astrophysics Source Code Library, ascl:2106.028.
\newblock \url{https://ui.adsabs.harvard.edu/abs/2021ascl.soft06028S}

\bibitem[{Spitler {et~al.}(2016)Spitler, Scholz, Hessels, Bogdanov, Brazier,
  Camilo, Chatterjee, Cordes, Crawford, Deneva, Ferdman, Freire, Kaspi,
  Lazarus, Lynch, Madsen, McLaughlin, Patel, Ransom, Seymour, Stairs, Stappers,
  van Leeuwen, \& Zhu}]{Spitler:2016tw}
Spitler, L.~G., Scholz, P., Hessels, J. W.~T., {et~al.} 2016, Nature, 531, 202,
  \dodoi{10.1038/nature17168}

\bibitem[{Tendulkar {et~al.}(2017)Tendulkar, Bassa, Cordes, Bower, Law,
  Chatterjee, Adams, Bogdanov, Burke-Spolaor, Butler, Demorest, Hessels, Kaspi,
  Lazio, Maddox, Marcote, McLaughlin, Paragi, Ransom, Scholz, Seymour, Spitler,
  van Langevelde, \& Wharton}]{Tendulkar:2017aa}
Tendulkar, S.~P., Bassa, C.~G., Cordes, J.~M., {et~al.} 2017, The Astrophysical
  Journal, 834, L7, \dodoi{10.3847/2041-8213/834/2/l7}

\bibitem[{Vikhlinin {et~al.}(2006)Vikhlinin, Kravtsov, Forman, Jones,
  Markevitch, Murray, \& Van~Speybroeck}]{Vikhlinin_2006}
Vikhlinin, A., Kravtsov, A., Forman, W., {et~al.} 2006, The Astrophysical
  Journal, 640, 691, \dodoi{10.1086/500288}

\bibitem[{Wechsler \& Tinker(2018)}]{Wechsler_2018}
Wechsler, R.~H., \& Tinker, J.~L. 2018, Annual Review of Astronomy and
  Astrophysics, 56, 435, \dodoi{10.1146/annurev-astro-081817-051756}

\bibitem[{Yao {et~al.}(2017)Yao, Manchester, \& Wang}]{Yao:2017aa}
Yao, J.~M., Manchester, R.~N., \& Wang, N. 2017, The Astrophysical Journal,
  835, 29, \dodoi{10.3847/1538-4357/835/1/29}

\bibitem[{Zhang {et~al.}(2021)Zhang, Yan, Li, Zhang, \& Wang}]{Zhang:2021aa}
Zhang, Z.~J., Yan, K., Li, C.~M., Zhang, G.~Q., \& Wang, F.~Y. 2021, The
  Astrophysical Journal, 906, 49, \dodoi{10.3847/1538-4357/abceb9}

\bibitem[{Zhou {et~al.}(2020{\natexlab{a}})Zhou, Newman, Mao, Meisner,
  Moustakas, Myers, Prakash, Zentner, Brooks, Duan, Landriau, Levi, Prada, \&
  Tarle}]{Zhou:2020ab}
Zhou, R., Newman, J.~A., Mao, Y.-Y., {et~al.} 2020{\natexlab{a}}, Monthly
  Notices of the Royal Astronomical Society, \dodoi{10.1093/mnras/staa3764}

\bibitem[{Zhou {et~al.}(2020{\natexlab{b}})Zhou, Newman, Dawson, Eisenstein,
  Brooks, Dey, Dey, Duan, Eftekharzadeh, Gazta{\~n}aga, Kehoe, Landriau, Levi,
  Licquia, Meisner, Moustakas, Myers, Palanque-Delabrouille, Poppett, Prada,
  Raichoor, Schlegel, Schubnell, Staten, Tarl{\'e}, \& Y{\`e}che}]{Zhou:2020ac}
Zhou, R., Newman, J.~A., Dawson, K.~S., {et~al.} 2020{\natexlab{b}}, Research
  Notes of the American Astronomical Society, 4, 181,
  \dodoi{10.3847/2515-5172/abc0f4}

\end{thebibliography}

\appendix
\section{Statistical errors on FRB locations}
\label{app:sec:localization}

Statistical errors in CHIME/FRB sky locations suppress the FRB-galaxy
power spectrum \clfg\ on small scales (large $\ell$).
The suppression takes the form \smash{$C_\ell^{fg} \rightarrow b_\ell C_\ell^{fg}$},
where $b_\ell$ is the ``beam'' transfer function.
Throughout the paper, we have modeled statistical errors as Gaussian,
which leads to a transfer function of the form \smash{$b_\ell = e^{-\ell^2/L^2}$}.

In this appendix, we will study statistical errors in more detail,
using toy models of the CHIME/FRB instrument and the FRB population.
Our conclusions are as follows:
\begin{itemize}
\item Statistical errors are not strictly Gaussian, but a Gaussian
 transfer function \smash{$b_\ell = e^{-\ell^2/L^2}$} is a good approximation
 within the error bars of our \clfg\ measurement.
\item Calculating $L$ from first principles is hard, since it depends
 on both the CHIME/FRB instrument and the FRB population.
 A plausible range of $L$-values is $315 \leq L \leq 1396$.
\end{itemize}
This justifies the methodology used throughout the paper, where a
Gaussian transfer function \smash{$b_\ell = e^{-\ell^2/L^2}$} is used, but
$L$ is a free parameter that we fit to the data, varying $L$ over the
range $315 \leq L \leq 1396$.

\subsection{Toy beam model 1: uniform density, center of nearest beam}
\label{app:ssec:selection_bias_neglected}

CHIME FRBs are detected by searching a $4\times 256$ regular
array of formed beams independently in real time.
A best-fit sky location is assigned to each detected FRB based
on the detection SNR (or nondetection) in each beam, using the
localization pipeline described by~\citet{Collaboration:2019tw,Collaboration:2021wz}.
For an FRB which is detected in a single beam, the localization
pipeline assigns sky location equal to the beam center.
For a multibeam detection, the assigned sky location is
roughly a weighted average of the beams where the event
was detected.

As a first attempt to model statistical errors in the localization
pipeline, suppose that when an FRB is detected we assign it to the
center of the closest FRB beam. This is a reasonable model for the
single-beam detections as described above.

We neglect wavelength dependence of the beam and evaluate at central
wavelength $\lambda = 0.5$ m.
We also neglect FRBs in sidelobes of the primary beam, since these
are a small fraction of the CHIME/FRB Catalog 1.
Finally, we assume that FRBs detected by CHIME/FRB are uniformly
distributed over the sky.
(This turns out to be a dubious approximation, as we will show in
the next subsection.)
What is $b_\ell$ in this toy model?

\begin{figure}
\centerline{\includegraphics[width=8.5cm]{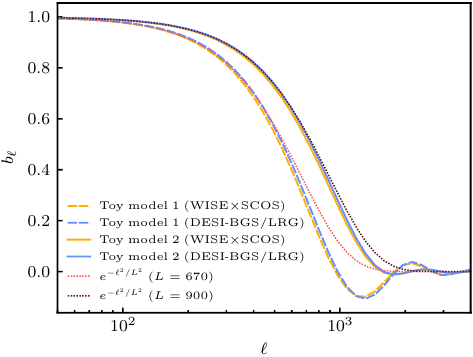}}
\caption{CHIME/FRB beam transfer function $b_\ell$ in a
toy beam model, without (model 1, \S\ref{app:ssec:selection_bias_neglected})
and with (model 2, \S\ref{app:ssec:selection_bias_included})
selection bias included.
Since $b_\ell$ is elevation dependent, the result is slightly different
after averaging over FRBs in the WISE$\times$SCOS (orange) and
DESI-BGS/LRG (blue) sky regions.
For values of $\ell$ which are resolved by the beam (say, $b_l \gtrsim 0.25$),
the beams are well approximated by Gaussians $b_\ell = e^{-\ell^2/L^2}$
(dotted curves).}
\label{fig:bl}
\end{figure}

Let $\Theta_e$ be the elevation of the detected FRB
(with the usual astronomical definition, i.e.~$\Theta_e=0$ for an FRB
on the horizon, or $\Theta_e=\pi/2$ for an FRB at zenith).
Let $\theta_x, \theta_y$ be east-west and north-south sky coordinates
in a coordinate system where the center of the formed beam is at $(0,0)$.
Let $S$ be the set of points closer to $(0,0)$ than any of the other
beam centers:
\be
S = \left[ -\frac{\theta_0}{2}, \frac{\theta_0}{2} \right] \times
\left[ -\frac{\theta_0}{2 \sin\Theta_e}, \frac{\theta_0}{2 \sin\Theta_e} \right]\, ,
\ee
where $\theta_0 = 23\farcm4$ in CHIME.
If the detected FRBs are uniformly distributed on the sky,
then the effective beam is
\be
b_\ell = \frac{\int_S d^2\theta \, J_0(\ell\theta)}{\int_S d^2\theta \, 1} \, ,
\ee
where $J_0(x)$ is a Bessel function.
For the CHIME/FRB catalog, which contains FRBs with different elevations $\Theta_e$,
we average $b_\ell$ over $\Theta_e$ values in the catalog.
It is straightforward to compute the elevation $\Theta_e$ for each FRB,
using values of RA, Dec, and time of observation taken directly from the catalog.
The resulting transfer function $b_\ell$ is shown in Figure~\ref{fig:bl},
and agrees well with a Gaussian transfer function \smash{$b_\ell = e^{-\ell^2/L^2}$}
with $L=670$.

\subsection{Toy beam model 2: including selection bias}
\label{app:ssec:selection_bias_included}

In the previous subsection, we neglected a selection bias:
an FRB is more likely to be detected if it is located at the
center of the beam (where the instrumental response is largest).
To account for this selection bias, we define the unnormalized intensity beam:
\be
B(\theta_x, \theta_y) = \frac{\sinc^2(\theta_x D_x /\lambda) \, \sinc^2((\theta_y D_y \sin\Theta_e)/\lambda)}{\sinc^2(\theta_x D_x / 4 \lambda)}\, ,
\ee
where $\theta_x, \theta_y, \Theta_e, \lambda$ are defined in \S\ref{app:ssec:selection_bias_neglected},
the CHIME aperture is modeled as a rectangle with dimensions $(D_x,D_y) = (80,100)$ meters,
and $\sinc(x) = \sin(\pi x) / (\pi x)$.

Assuming a Euclidean FRB fluence distribution \smash{$N(\ge F) \propto F^{-3/2}$}
(consistent with statistical analysis of the CHIME/FRB Catalog 1~\citep{Collaboration:2021wz}),
the probability of detecting an FRB at sky location $(\theta_x, \theta_y)$ is
$\propto B(\theta_x, \theta_y)^{3/2}$.
Therefore, the beam transfer function is
\be
b_\ell = \frac{\int_S d^2\theta \, B(\theta)^{3/2} J_0(\ell\theta)}{\int_S d^2\theta \, B(\theta)^{3/2}}
\ee
averaged over catalog elevations $\Theta_e$ as in the previous subsection.
The resulting transfer function $b_\ell$ is shown in Figure~\ref{fig:bl}
and agrees well with a Gaussian transfer function \smash{$b_\ell = e^{-\ell^2/L^2}$}
with $L=900$.

\subsection{Plausible range of $L$-values}

Comparing the last two subsections, we see that the selection
bias considered in~\S\ref{app:ssec:selection_bias_included} increases 
the effective value of $L$ by 34\%.
This treatment of selection bias is incomplete, and a full study
is outside the scope of this paper.
For example, $b_\ell$ depends on wavelength $\lambda$, so there
is a selection bias involving FRB frequency spectra.
In addition, we have not attempted to model multibeam detections,
which will be better localized than single-beam detections.
Given these sources of modeling uncertainty, rather than
trying to model the value of $L$ precisely, we will assign
a range of plausible $L$-values.

To assign a smallest plausible $L$-value, we make assumptions
that lead to the largest plausible localization errors.
We start with the toy beam model $b_\ell$ from~\S\ref{app:ssec:selection_bias_included},
with $\lambda=0.75$\,m (the longest wavelength in CHIME).
We then convolve with a halo profile 
\citep[\smash{$b_\ell \rightarrow b_\ell u_\ell(M,z)^2$},
where $u_\ell(M,z)$ is a Navarro-Frenk-White (NFW) density profile;][]{Navarro:1996gj},
taking the halo mass $M$ to be large ($M=10^{14.5}~h^{-1}M_\odot$)
and the redshift to be small ($z=0.05$).
These specific values are somewhat arbitrary, but the goal
is to establish a baseline plausible value of $L_{\rm min}$,
not model a precise value of $L$.
With the assumptions in this paragraph, we get $L_{\rm min}=315$.

Similarly, to assign a largest plausible $L$-value, we make assumptions
that lead to the smallest plausible localization errors.
We use the smallest toy model from~\S\ref{app:ssec:selection_bias_included}
with $\lambda=0.375$\,m (the shortest wavelength in CHIME).
We assume that 40\% of the events are multibeam detections,
and that multibeam detections have localization errors that
are smaller by a factor 3.
As in the previous section, these specific values are somewhat arbitrary,
but the goal is to establish a baseline plausible value of $L_{\rm max}$,
not model a precise value of $L$.
With the assumptions in this paragraph, we get $L_{\rm max}=1396$.

\section{Null tests}
\label{app:sec:null_tests}

As a general check for robustness of our FRB-galaxy correlation \clfg,
we would like to check that \clfg\ does not depend on external variables,
for example time of day (TOD).
Our methodology for doing this is as follows.
We divide the FRB catalog into low-TOD and high-TOD subcatalogs, cross-correlate
each subcatalog with a galaxy sample, and compute the difference power spectrum:
\be
d\hC_{\ell}^{fg} \equiv \hC_{\ell}^{fg\,({\rm low})} - \hC_{\ell}^{fg\,({\rm high})} \, . \label{eq:dclfgli}
\ee
Recall that for a non-null power spectrum $\hC_\ell$, we compressed the $\ell$-dependence
into a scalar summary statistic $\hat\alpha_L$ by taking a weighted $\ell$-average (Eq.~\ref{eq:alpha-hat}).
Analogously, we compress the difference spectrum \smash{$d\hC_\ell^{fg}$} into a summary
statistic $\hat\beta_L$, defined by
\be
\hat\beta_{L} = \sum_{\ell\ge \ell_{\rm min}} (2\ell+1) \frac{e^{-\ell^2/L^2}}{C_\ell^{gg}} d\hC_{\ell}^{fg} \, , \label{eq:beta-hat}
\ee
where $L$ is an angular scale parameter.
Next, by analogy with $\SNR_L$ (defined previously in Eq.~\ref{eq:sigma-loc}), we define
\be
\Delta_{L} = \frac{\hat\beta_{L}}{\Var(\hat\beta_{L})^{1/2}} \, . \label{eq:Delta-loc}
\ee
The value of $\Delta_L$ quantifies consistency (in ``sigmas'')
between \clfg\ for the low-TOD and high-TOD subcatalogs.

We fix $L=1000$, and consider three choices of galaxy catalog:
WISE$\times$SCOS with $z \ge 0.3125$, DESI-BGS with $z \ge 0.295$,
and DESI-LRG with $z \le 0.485$.
These redshift ranges are ``cherry-picked'' to
maximize the FRB-galaxy cross-correlation (see Figure~\ref{fig:clfg_sl_binned}),
but this cherry-picking should not bias the difference statistic $\Delta_L$.
With these choices, we find $\Delta_L = \{ 1.22, -0.21, 1.30 \}$
for WISE$\times$SCOS, DESI-BGS, and DESI-LRG respectively.
Therefore, there is no statistical evidence for dependence of \clfg\
on time of day, since a 1.22$\sigma$, 0.21$\sigma$, or 1.30$\sigma$
result is not statistically significant.

This test can be generalized by splitting on a variety of external
variables (besides TOD).
In Table~\ref{tab:null_tests}, we identify 12 such variables
and denote the corresponding $\Delta_L$ values (with $L=1000$)
by $\Delta_i$, where $i \in \{1, 2, 3, \dots, 12\}$.
We note that these 12 tests are nonindependent, for example SNR is
correlated with fluence.
We also note that for many of these tests detection of a nonzero
difference spectrum \smash{$d\hC_\ell^{fg}$} does not necessarily indicate
a problem.
For example, DM dependence of \clfg\ is expected at some level,
since \clfg\ is redshift dependent, and DM is correlated with redshift.

\begin{table*}
\begin{center}
\begin{tabular}{@{\hskip 0.5cm}c@{\hskip 0.6cm}|@{\hskip 0.6cm}c@{\hskip 0.6cm}c@{\hskip 0.6cm}|
	@{\hskip 0.6cm}c@{\hskip 0.6cm}cc}
\hline\hline
    Parameter & Median & $\Delta_{i}$ & Median & \multicolumn{2}{c}{$\Delta_{i}$} \\
    & (WISE$\times$SCOS)  &{WISE$\times$SCOS} & (DESI) & DESI-BGS & DESI-LRG \\
\hline
    DM [\pcc] & 535.08 & 0.33 & 536.41 & $-1.68$ & $-0.95$ \\
    SNR & 20.2 & 0.58 & 20.2 & $0.00$ & 0.42 \\
    Scattering time [ms]& 1.331 & 1.49 & 1.423 & 0.93 & 0.38 \\
    Pulse width [ms] & 0.988 & 0.59 & 1.052 & 0.99 & 0.24 \\
    Spectral index & 2.866 & 0.68 & 2.075 & 0.76 & $-0.25$ \\
    Fluence [Jy\,ms]& 3.503 & 1.28 & 3.115 & $-1.00$ & 2.16 \\
    Bandwidth [MHz]& 332.09 & $-0.44$ & 358.09 & 0.80 & 1.58 \\
    Galactic $|b|$ & 38\fdg26 & 0.59 & 38\fdg24 & $-1.27$ & $-1.95$ \\
    Catalog localization error & 10\farcm12 & 0.52 & 9\farcm53 & 2.16 & 1.19 \\
    ${\rm TOA} - 58528$ [MJD] & 0.3686595 & 0.99 & 4.8473498 & 1.16 & 0.63 \\
    Peak frequency [MHz]& 463.525 & $-0.63$ & 449.036 & 1.97 & 1.30 \\
    Time of day [hr]& 9.887 & 1.22 & 10.132 & $-0.21$ & 1.30 \\
\hline\hline
\end{tabular}
    \caption{\label{tab:null_tests}
Null tests in Appendix~\ref{app:sec:null_tests}.
For each parameter, we split the FRB catalog into ``low'' and ``high''
subcatalogs by comparing the parameter value to its median.
(The median value is slightly different for FRBs in the WISE$\times$SCOS
and DESI footprints.)
We correlate both subcatalogs with the galaxy surveys, and compute
the statistic $\Delta = (\Delta_L)_{L=1000}$ (defined in Eq.~\ref{eq:Delta-loc}), which
measures consistency of the FRB-galaxy correlation in ``sigmas''.}
\end{center}
\end{table*}

There are a few $\sim$2$\sigma$ outliers in Table~\ref{tab:null_tests},
but a few outliers are unsurprising, so it is not immediately clear
whether the $\Delta_i$ values in Table~\ref{tab:null_tests} are
statistically different from zero.
To answer this question, we reduce the 12-component vector $\Delta_i$
into a scalar summary statistic, in a few different ways as follows.

Our first summary statistic is intended to test whether the most anomalous
$\Delta_i$-value in each column of Table~\ref{tab:null_tests} is statistically
significant. We define
\be
\Delta_{\rm max} = \max_i |\Delta_{i}| \, .  \label{eq:delta_max_def}
\ee
We then compare these values of $\Delta_{\rm max}$ to an ensemble of mocks.
The mocks are constructed by randomizing the RA of each FRB in the catalog,
keeping all other FRB properties (DM, SNR, etc.) fixed.
This preserves any correlations which may be present between FRB properties.
In the special case of the $|b| \ge 17^\circ$ null test, we recompute the value
of $b$ after randomizing RA.

In Table~\ref{tab:null_tests_glo}, we report the \pvalue for
each $\Delta_{\rm max}$, i.e.~the fraction of mocks whose $\Delta_{\rm max}$
exceeds the ``data'' value.
No statistically significant deviation from $\Delta_{\rm max}=0$ is seen.

Our second summary statistic is intended to test whether the 12-component
vector $\Delta_i$ is consistent with a multivariate Gaussian distribution.
We define:
\be
\chi^2 = \sum_{i,i'} \Delta_{i} \, \Cov\left(\Delta_{i}, \Delta_{i'}\right)^{-1} \Delta_{i'} \, ,  \label{eq:chi2_def}
\ee
where the covariance $\Cov(\Delta_{i}, \Delta_{i'})$ is estimated from mock
FRB catalogs, constructed as described above.

As before, to assign statistical significance, we compare the ``data'' value of 
$\chi^2$ to an ensemble of mocks 
and report the associated \pvalue in Table~\ref{tab:null_tests_glo}.
We find borderline evidence for $\chi^2 \ne 0$ for DESI-BGS ($p=0.030$),
but interpret this as inconclusive, since Table~\ref{tab:null_tests_glo}
contains six \pvalues, so one \pvalue as small as 0.03 is unsurprising
(this happens with probability $\approx$0.18).

\begin{table*}
\begin{center}
\begin{tabular}{@{\hskip 0.6cm}c@{\hskip 0.65cm}|@{\hskip 0.65cm}c@{\hskip 0.5cm}c@{\hskip 0.65cm}|
	  @{\hskip 0.65cm}c@{\hskip 0.5cm}c@{\hskip 0.65cm}@{\vrule width 0.9pt}@{\hskip 0.65cm}c@{\hskip 0.5cm}c@{\hskip 0.6cm}}
    \hline\hline
    Galaxy sample & $\Delta_{\rm max}$ & \pvalue & $\chi^2$ & \pvalue & KS \pvalue  & AD \pvalue \\
\hline
     WISE${\times}$SCOS & 1.49 & 0.779 & 9.26 & 0.659 & 0.037 & 0.067 \\
     DESI-BGS & 2.16 & 0.270 & 22.96 & 0.030 & 0.381 & 0.250 \\
     DESI-LRG & 2.16 & 0.274 & 17.26 & 0.145 & 0.113 & 0.171 \\
\hline\hline
\end{tabular}
    \caption{\label{tab:null_tests_glo}
Summary statistics for the 12 null tests in Table~\ref{tab:null_tests}.
As described in Appendix~\ref{app:sec:null_tests}, we reduce the 12-component
vector $\Delta_i$ into two scalar summary statistics $\Delta_{\rm max}, \chi^2$,
shown in the first four columns along with associated \pvalues from an ensemble
of mocks.
The last two columns compare the $\Delta_i$ values for each galaxy sample
to a ``jackknife'' ensemble defined by randomly splitting the CHIME/FRB catalog.}
\end{center}
\end{table*}

\begin{figure}
\centerline{\includegraphics[width=8.5cm]{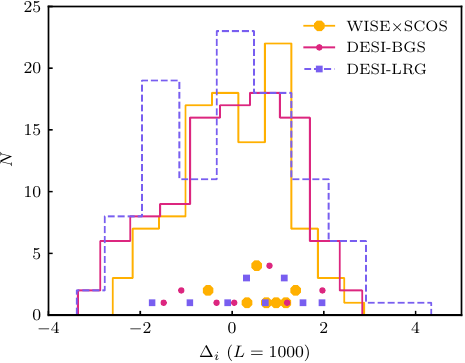}}
    \caption{Histograms of the statistic $\Delta_i$ for the 12 null tests (filled markers) and 100 jackknives (lines).
   Using the KS and AD tests, the distributions are found to be consistent (Appendix~\ref{app:sec:null_tests}).}
\label{fig:hist_null_tests}
\end{figure}

Finally, we compare the set of 12 $\Delta_i$ values to a jackknife
distribution, obtained by randomly splitting the FRB catalog in half.
We do this comparison using
the 2-sample Kolmogorov-Smirnov \citep[KS,][]{Hodges:1958aa} and Anderson-Darling \citep[AD,][]{Scholz:1987aa} tests.
Figure~\ref{fig:hist_null_tests} compares the two distributions for the three galaxy samples,
and the last two columns of Table~\ref{tab:null_tests_glo} summarize our results.
As in the previous paragraph, there is one outlier: the WISE$\times$SCOS KS \pvalue is 0.037,
which we interpret as inconclusive, since it is one out of six \pvalues in the table
(as in the previous paragraph).

Summarizing this appendix, we do not find statistically significant evidence that the
FRB-galaxy clustering signal studied in this paper depends on any of
the parameters in Table~\ref{tab:null_tests}.

\section{Tail-fitting procedure}
\label{app:tail_fitting}

In~\S\ref{ssec:redshift_dependence}, we assign statistical significance
of the FRB-galaxy detection, by defining a frequentist statistic $\SNR_{\rm max}$,
and ranking the ``data'' value \sdx\ within a histogram
of simulated values \smx.
This procedure is conceptually straightforward, but there is a technical
challenge: because \sdx\ turns out to be an extreme
outlier, a brute-force approach requires an impractical
number of simulations.
Therefore, we fit the tail of the \smx\ distribution
to an analytic distribution and assign statistical significance (or \pvalue)
analytically.

\begin{figure}
\centerline{\includegraphics[width=8.5cm]{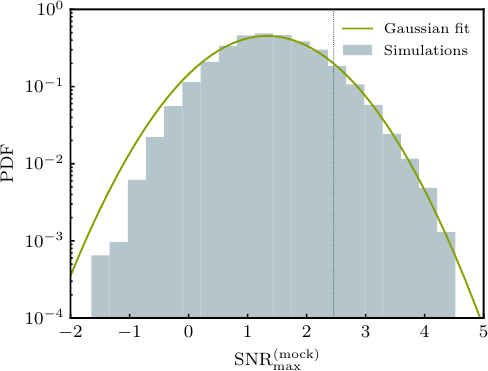}}
\caption{Gaussian fit to the tail of the \smx\
 distribution from Appendix~\ref{app:tail_fitting}. For the top $\sim$10\%
 of the samples (i.e.~to the right of the dotted line) the agreement between
 the fit and the simulations is excellent. This plot is for WISE$\times$SCOS;
 the other two cases (DESI-BGS, DESI-LRG) are similar.}
\label{fig:tail_fit}
\end{figure}

Empirically, we find that the top 10\% of the \smx\
distribution agrees well with the top 10\% of a Gaussian distribution, as
shown in Figure~\ref{fig:tail_fit}.
The parameters of the Gaussian distribution were determined as follows.
Let $p(x|\mu,\sigma)$ denote a Gaussian distribution with mean $\mu$
and variance $\sigma^2$:
\be
p(x|\mu,\sigma) = \frac{1}{\sigma\sqrt{2\pi}} e^{-(x-\mu)^2/(2\sigma^2)} \, .
\ee
Let \smash{$\Sigma_+ \subset \mathbb{R}$} be the top 10\% of the simulated
\smx\ values, and let $\Sigma_-$ be the bottom 90\%.
Let $\Sigma_0 \in \mathbb{R}$ be the 90th percentile of the 
\smx\ distribution.
Then, we choose parameters $(\mu,\sigma)$ to maximize the likelihood
function:
\begin{align}
\log \mathcal{L}(x | \mu,\sigma) 
  &= \bigg( \sum_{x\in \Sigma_+} \log p(x|\mu,\sigma) \bigg) \nn \\
  & \hspace{0.94cm} + \big| \Sigma_- \big| \log \int_{-\infty}^{\Sigma_0} p(x|\mu,\sigma) \, ,
\end{align}
where $x$ denotes mock realizations.
This likelihood function has been constructed to fit parameters
to the details of the $\Sigma_+$ values, while putting all $\Sigma_-$
values into a single coarse bin.

Figure~\ref{fig:tail_fit} is a good visual test for goodness of fit,
but as a more quantitative test, we compare the upper 10\% of the simulated
histogram with the Gaussian fit using a KS test.
We find that the two distributions agree to $1\sigma$ (and likewise for
the other two cases, DESI-BGS and DESI-LRG).

In Table~\ref{tab:tail_fits}, we compute statistical significance
for each of the three surveys, in two different ways.
The ``brute-force'' \pvalue is obtained by counting the number
of simulated \smx\ values (out of $10^4$
total simulations) that exceed \sdx.
The ``analytic'' \pvalue is obtained by fitting the top 10\%
of the simulated \smx\ values to a Gaussian
distribution, as described above, and evaluating the CDF of the
distribution at \sdx.
The brute-force values are either uninformative (for
WISE$\times$SCOS), or have large Poisson uncertainties (for the
other two surveys), so we have quoted the analytic \pvalues as
our ``bottom-line'' detection significances throughout the paper.

\begin{table}
\begin{center}
\begin{tabular}{@{\hskip 0.2cm}c@{\hskip 0.75cm}c@{\hskip 0.8cm}c@{\hskip 0.3cm}}
\hline\hline
 Survey & Brute-force & Analytic \\ \hline
 WISE$\times$SCOS & 0/10000 & $2.7 \times 10^{-5}$ \\
 DESI-BGS & 4/10000 & $3.1 \times 10^{-4}$ \\
 DESI-LRG & 5/10000 & $4.1 \times 10^{-4}$ \\
\hline\hline
\end{tabular}
\end{center}
\caption{``Brute-force'' and analytic \pvalues, computed as described
 in Appendix~\ref{app:tail_fitting}.}
\label{tab:tail_fits}
\end{table}

\end{document}